\documentclass[acs,jctc,preprint,floatfix]{revtex4-1}

\usepackage{graphicx}   
\usepackage{hyperref}   
\usepackage{amsmath}    
\usepackage{algorithm}
\usepackage{algpseudocode}
\usepackage{float}
\usepackage{tabularx}
\usepackage{array}      
\usepackage{xcolor}     
\usepackage[bottom]{footmisc}     

\begin{document}

\title{Tensor Hypercontraction Form of the Perturbative Triples Energy in Coupled-Cluster Theory}

\author{Andy Jiang}
\affiliation{Center for Computational Quantum Chemistry, Department of Chemistry, University of Georgia, Athens, GA 30602}
\affiliation{Center for Computational Molecular Science and
Technology, School of Chemistry and Biochemistry, School of Computational
Science and Engineering, Georgia Institute of Technology, Atlanta, GA
30332-0400}
\author{Justin M. Turney}
\email{justin.turney@uga.edu}
\affiliation{Center for Computational Quantum Chemistry, Department of Chemistry, University of Georgia, Athens, GA 30602}

\author {Henry F. Schaefer III}
\email{ccq@uga.edu}
\affiliation{Center for Computational Quantum Chemistry, Department of Chemistry, University of Georgia, Athens, GA 30602}

\date{\today}

\begin{abstract}

We present the working equations for a reduced-scaling
method of evaluating the perturbative triples (T) energy in coupled-cluster
theory, through the tensor hypercontraction (THC) of the triples amplitudes ($t_{ijk}^{abc}$). Through our method we can reduce the scaling of the (T) energy
from the traditional $\mathcal{O}(N^{7})$ to a more modest $\mathcal{O}(N^{5})$. We also
discuss implementation details to aid future research, development, and
software realization of this method. Additionally, we show that this method yields sub-millihartree (mEh) differences from CCSD(T)
when evaluating absolute energies, and sub-0.1 kcal/mol energy differences when evaluating relative energies. Finally, we
demonstrate that this method converges to the true
CCSD(T) energy through the systematic increasing of the
rank or eigenvalue tolerance of the orthogonal projector, as well
as exhibiting sub-linear to linear error growth with respect to
system size.

\end{abstract}

\maketitle

\section{Introduction}

Coupled-cluster (CC) theory \cite{Crawford2007,Bartlett2007} is one of the most important advances of modern
quantum chemistry, allowing for a polynomial-time evaluation of the
electronic energies and wavefunction of a molecule, as a size-extensive
alternative to truncated configuration interaction (CI) methods \cite{Cramer2002,Sherrill1999}. Truncated CC
methods also avoid the intractable super-exponential scaling of full
configuration interaction (FCI), yielding reasonable and chemically accurate
relative energies compared to both the FCI limit and to
experimental results, especially in the context of CCSD(T), also known as the ``gold standard'' method in computational quantum
chemistry \cite{Raghavachari1989}.
The tractability and accuracy of CC methods make the development of efficient CC methods crucial for the future of quantum
chemistry, as evaluation of accurate energies and wavefunctions is made
possible for larger and more complex systems through hardware advances such as massively parallel computing
\cite{Hirata2003,Auer2006,Janowski2007,Janowski2008,vanDam2011,Deumens2011,Kobayashi1997,Anisimov2014,Solomonik2014,Peng2016,Lyakh2019,Gyevi2020,Peng2020,Datta2021,Gyevi2021,Kowalski2021,Calvin2021}
and GPUs \cite{Seritan2020,Wang2020,Peng2019,Kaliman2017,DePrince2014,Ma2011,DePrince2011}.

However, there is still a tremendous gap in applicability between
coupled-cluster theories (formally scaling at least $\mathcal{O}(N^{6})$) and lower-scaling
methods like M{\o}ller--Plesset perturbation theory (MP2) \cite{Moller1934,Cremer2011} and density
functional theory (DFT) \cite{Hohenberg1964,Kohn1965} (scaling $\mathcal{O}(N^{5})$ or better). Because of this,
DFT and MP2 can be run on system tens or even hundreds of times the
size of a system typically evaluated with CC methods \cite{Maurer2014,Dawson2022}. To close the gap
between CC and less reliable electron correlation methods, it is useful
to devise approximation schemes to CC which reduce the scaling, but
also allow a means
to systematically control the error compared to the non-approximated CC method. One such approach involves local-correlation
\cite{Li2002,Li2006,Li2009,Neese2009,Li2010a,Li2010b,Rolik2011,Rolik2013,Riplinger2013a,Riplinger2013b,Liakos2015,Schwilk2017}, such as used in the
DLPNO methods \cite{Pinski2015,Riplinger2016}. With large enough
molecules, these methods achieve asymptotic linear-scaling.

Another approach is the rank reduction of the coupled-cluster
amplitudes \cite{Parrish2019}, using orthogonal projectors that transform the single and double cluster-amplitudes into a smaller basis
\begin{align}
    T^{V} &= U^{V}_{ia}t_{i}^{a}\label{eq:1} \\
    T^{VW} &= U^{V}_{ia}t_{ij}^{ab}U^{W}_{jb} \ . \label{eq:2}
\end{align}
Because of the orthogonal nature of the projectors, getting the full amplitudes from the
rank-reduced form is trivial
\begin{align}
    t_{i}^{a} &= U^{V}_{ia}T^{V}\label{eq:3} \\
    t_{ij}^{ab} &= U^{V}_{ia}T^{VW}U^{W}_{jb} \ . \label{eq:4}
\end{align}

As shown by Parrish and co-workers, the size of the $V$ and $W$ indices, also known as the projector rank, can be made directly proportional to the
system size, while maintaining a set relative error from the absolute energy of a molecule \cite{Parrish2019}. More recently, Hohenstein et. al. have
shown how to create a tensor hypercontracted (THC) form of the $t_{ij}^{ab}$ amplitudes,
through the CANCENCOMP/PARAFAC (CP) decomposition \cite{Kolda2009} of the orthogonal projectors \cite{Hohenstein2022}.
\begin{align}
    U^{V}_{ia} &= \sum_{X} y_{i}^{X}y_{a}^{X}\tau_{VX} \label{eq:5}\\
    t_{ij}^{ab} &= \sum_{XY} y_{i}^{X}y_{a}^{X}\widetilde{T}^{XY}y_{j}^{Y}y_{b}^{Y} \label{eq:6} \\
    \widetilde{T}^{XY} &= \sum_{VW} \tau_{VX} T^{VW} \tau_{WY} \label{eq:7}
\end{align}
Parrish and Hohenstein have also shown that, in the context of CCSD, the size of the $X$
index can be made proportional to the system size to maintain a set relative error. Rank-reduction methods have also been applied to coupled-cluster theories involving
higher levels of excitation, recently by Lesiuk with the SVD-CCSDT method \cite{Lesiuk2019}, where the concept of orthogonal projectors is used to approximate the triples amplitude in CCSDT theory
\begin{equation}
    t_{ijk}^{abc} = U^{U}_{ia}U^{V}_{jb}U^{W}_{kc}T^{UVW} \label{eq:8} \ .
\end{equation}

In the following sections, we will combine the concepts of orthogonal projectors and THC to develop working equations for a reduced-scaling variant of the non-iterative
perturbative triples correction to the CCSD energy \cite{Raghavachari1989}. Recently, Lesiuk derived an
$O(N^{6})$ approach to the (T) energy with orthogonal
projectors which he calls RR-CCSD(T) \cite{Lesiuk2022}. In the current paper, we will improve upon the work of Lesiuk's
approach utilizing tensor hypercontraction. Similar to how the THC-CCSD method \cite{Hohenstein2022} improves
upon the RR-CCSD method \cite{Parrish2019,Lesiuk2022}, our new approach, which we name THC-CCSD(T), will
commensurately enhance RR-CCSD(T), reducing the scaling of Lesiuk's from $\mathcal{O}(N^6)$ to $\mathcal{O}(N^5)$. For consistency, we will use
many of the same formalisms as Lesiuk \cite{Lesiuk2019} and Hohenstein \cite{Hohenstein2022}.

\section{Theory}

\subsection{Notation}

We will use the following conventions to describe the
indices appearing in this work:
\begin{itemize}
    \item $i, j, k, l$: Occupied molecular orbitals, which ranges from 1 to $n_{occ}$.
    \item $a, b, c, d$: Virtual molecular orbitals, which ranges from 1 to $n_{virt}$.
    \item $P, Q$: Auxiliary indices of density-fitted/Cholesky-decomposed ERIs, which ranges from 1 to $n_{aux}$.
    \item $w, v$: Laplace denominator weight indices, which ranges from 1 to $n_{w}$.
    \item $U, V, W$: Rank-reduced dimension of the doubles orthogonal projector, which ranges from 1 to $n_{proj}$.
    \item $A, B, C$: Rank-reduced dimension of the triples
    orthogonal projector, which ranges from 1 to $n_{proj}$.
    \item $X, Y, Z$: CP-decomposition rank of the triples orthogonal projector, which ranges from 1 to $n_{proj}$.
\end{itemize}
The relative sizes of the indices are as follows:
\begin{equation}
    n_{occ} < n_{virt} < n_{aux} \approx n_{proj}
\end{equation}
Note that $n_{w}$ does not grow with increasing molecular system size, and therefore, run-time
analysis of intermediates with $w, v$ indices will only treat the Laplace index as a prefactor.

The frozen-core approximation was used in all post-Hartree--Fock computations in this work; i.e., the 1$s$ electrons are not correlated for all first-row atoms. The occupied space $n_{occ}$ always refers to the number of correlated occupied orbitals.
Einstein summation convention is used throughout -- all indices appearing on the right-hand side but not on the left-hand side of an expression are summed over.

\subsection{Perturbative Triples Correction to CCSD}

CCSD is often not sufficient to obtain ``chemically-reliable'' theoretical predictions, and it has been shown that only after triple excitations are considered that relative energies of under 1 kcal/mol can be regularly achieved
\cite{Riley2010,Karton2006,Tajti2004,Bak2000,Hopkins2004,Bartlett1990}. However, an explicit treatment
of all triples has a very high cost of $\mathcal{O}(N^{8})$.
Therefore, the triples amplitudes are often determined in a perturbative
manner, based on the work of Raghavachari and co-workers \cite{Raghavachari1989}.
In their formalism, the perturbative
triples correction to the CCSD energy is defined as
\begin{equation}
    E^{(T)} = E^{[4]}_{T} + E^{[5]}_{ST}
\end{equation}
where
\begin{align}
    E^{[4]}_{T} &= \left<T_{2} \mid [W, T_{3}]\right> \\
    E^{[5]}_{ST} &= \left< T_{1} \mid [W, T_{3}] \right> \ .
\end{align}
$T_{1}$, $T_{2}$, and $T_{3}$ are known as the ``cluster operators'' and, in second-quantization formalism, are defined as
\begin{align}
    T_{1} &= t_{i}^{a}E_{ai}\\
    T_{2} &= t_{ij}^{ab}E_{ai}E_{bj}\\
    T_{3} &= t_{ijk}^{abc}E_{ai}E_{bj}E_{ck}
\end{align}
$E_{ai}$ represents the singlet, spin-adapted excitation operator, and
is defined as
\begin{equation}
    E_{ai} = a^{\dagger}_{a} a_{i} +
                {\bar a}^{\dagger}_{a} {\bar a}_{i}
\end{equation}
where the barred creation/annihilation operators refer to the beta spin orbitals and nonbarred refer to the alpha spin orbitals.

The accuracy of the (T) method stems from a highly favorable error cancellation between $E^{[4]}_{T}$ and $E^{[5]}_{ST}$.
In restricted, single-reference, closed-shell coupled cluster theory, one can write the equation for the $(T)$ correction as \cite{Gyevi2019}

\begin{equation}\label{eq:pTenergy}
    E^{(T)} = \frac{1}{3} 
    \frac{(4W_{ijk}^{abc} + W_{ijk}^{bca} + W_{ijk}^{cab})(V_{ijk}^{abc} - V_{ijk}^{cba})}{\epsilon_{i} + \epsilon_{j} + \epsilon_{k} - \epsilon_{a} - \epsilon_{b} - \epsilon_{c}}
\end{equation}
where
\begin{equation}\label{eq:Wijkabc}
    W_{ijk}^{abc} = P_{L}\left[
    (ia|bd)t_{kj}^{cd} - 
    (ia|jl)t_{kl}^{cb}\right]
\end{equation}
and
\begin{equation}\label{eq:Vijkabc}
    V_{ijk}^{abc} = W_{ijk}^{abc} + P_{S}\left[t_{i}^{a}(jb|kc)\right]
\end{equation}
Following the formalism of Lesiuk \cite{Lesiuk2019}, we define $P_{L}$ and $P_{S}$, or the ``long'' and ``short'' permutation
operations as
\begin{align}
    P_{L}(A_{ijk}^{abc}) &= A_{ijk}^{abc} + A_{ikj}^{acb} + A_{jik}^{bac} + A_{jki}^{bca}
                            + A_{kij}^{cab} + A_{kji}^{cba}\\
    P_{S}(A_{ijk}^{abc}) &= A_{ijk}^{abc} + A_{jik}^{bac} + A_{kij}^{cab}
\end{align}

The perturbative triples amplitude ($t_{ijk}^{abc}$), is defined as

\begin{equation}\label{eq:pTamplitude}
    t_{ijk}^{abc} = \frac{W_{ijk}^{abc}}{\epsilon_{i} + \epsilon_{j} + \epsilon_{k} - \epsilon_{a} - \epsilon_{b} - \epsilon_{c}}
\end{equation}

Using the perturbative triples amplitude, as well as the permutational symmetry of the Laplace denominator, one can rewrite
Equation \ref{eq:pTenergy} as:

\begin{equation}\label{eq:pTenergynew}
    E^{(T)} = t_{ijk}^{abc} \cdot (\frac{4}{3}V_{ijk}^{abc}
        - 2V_{ijk}^{cba} + \frac{2}{3}V_{ijk}^{cab})
\end{equation}

\noindent We will use this equation when deriving the formulas for the
THC-CCSD(T) energy. \\

The cost of evaluating expression \ref{eq:pTenergynew} scales as $\mathcal{O}(N^{6})$. However, the cost of evaluating
expression \ref{eq:Wijkabc} scales as $\mathcal{O}(N^{7})$,
leading to an overall unfavorable $\mathcal{O}(N^{7})$ scaling of the CCSD(T) method.

\subsection{Orthogonal Projectors}

One crucial step of rank-reduced coupled cluster methods is the the formation of the orthogonal
projectors to reduce the dimensionality of the amplitudes, as given in equations \ref{eq:1}-\ref{eq:4} and \ref{eq:8}.
There are a variety of methods that can be used to compute orthogonal projectors. One such method for the CCSD doubles
amplitude is to form them from the definition of the MP2 $t_{ij}^{ab}$ amplitudes \cite{Parrish2019}.
\begin{equation}
    t_{ij}^{ab} = \frac{(ia|jb)}{\epsilon_{i} + \epsilon_{j} - \epsilon_{a} - \epsilon_{b}}
\end{equation}
Using density-fitting (DF) \cite{Dunlap1979,Weigend1998}, also known as resolution-of-the-identity (RI), or Cholesky Decomposition (CD) \cite{Roeggen2008},
the set of electron-repulsion integrals (ERIs) in the molecular orbital (MO) basis $(ia|jb)$ can be
written as follows \cite{DePrince2013}:
\begin{equation}\label{eq:df-ri-approx}
    (ia|jb) \approx 
    B^{Q}_{ia} B^{Q}_{jb}
\end{equation}

\noindent The energy denominator can
be factored with a constant-sized index $w$ (with growing molecular system size) through the Laplace denominator approach \cite{Haser1992}
\begin{equation}
    \frac{1}{\epsilon_{i} + \epsilon_{j} - \epsilon_{a} - \epsilon_{b}} = -
    D_{iw}D_{jw}D_{aw}D_{bw} \label{eq:24}
\end{equation}

\noindent Combining these techniques, and the following intermediates, as defined by Parrish and co-workers \cite{Parrish2019},
\begin{align}
    L^{Qw}_{ia} &= D_{iw}D_{aw} B^{Q}_{ia} \\
    M_{Pw,Qv} &= 
    L^{Pw}_{ia} L^{Qv}_{ia}
\end{align}
allows us to diagonalize $M$ and form the MP2 projector ($U^V_{ia}$) as
\begin{align}
    M_{Pw,Qv} &= V^{V}_{Pw}\tau^{V}V^{V}_{Qv} \label{eq:29} \\
    U^{V}_{ia} &= L^{Qw}_{ia}V^{V}_{Qw}\frac{1}{\sqrt{\tau^{V}}} \ .
\end{align}
Note that the size of the index V can be truncated based on the magnitude of the corresponding eigenvalue
$\tau^{V}$. Even though the diagonalization of $M$ is technically cubic-scaling, the size of the $w$ index
can provide a large prefactor. In the case of larger molecules, the size of the $V$ index is often much
smaller than the size of the $[Qw]$ index, and thus truncated diagonalization approaches like the one given
in reference \citenum{Halko2011} may be used. Overall, this approach scales $\mathcal{O}(N^{4})$. Similarly, projectors can be derived from
MP3, albeit the equations are more complex \cite{Parrish2019,Lesiuk2022},

For triples amplitudes, we present two approaches devised by Lesiuk. In his SVD-CCSDT algorithm \cite{Lesiuk2019}, he took guess $t_{ijk}^{abc}$ amplitudes, such
as from CC3, and applied either a TUCKER-3 decomposition (scaling $\mathcal{O}(N^{8})$) or an iterative SVD approach
(scaling $\mathcal{O}(N^{6})$), yielding the form of
equation \ref{eq:8}.

In his RR-CCSD(T) paper, Lesiuk devised an $\mathcal{O}(N^{5})$ scheme to compute projectors from the form of the perturbative triples
amplitudes (Equation \ref{eq:pTamplitude}), in a variant of HO-OI (Higher Order-Orthogonal Iteration) \cite{Lesiuk2022}. The steps of
the algorithm are as follows:

\begin{itemize}
    \item Start with the a guess of the triples projector $V_{ia}^{A}$. This can be done naively by setting $V_{ia}^{A} =
    U_{ia}^{A}$ from the doubles amplitudes.
    \item Evaluate $t_{ia,BC}$ from the current guess of the triples amplitudes, where
    \begin{equation}
        t_{ia,BC} = t_{ijk}^{abc}V_{jb}^{B}V_{kc}^{C}
    \end{equation}
    By using the explicit expression for $t_{ijk}^{abc}$ and
    $W_{ijk}^{abc}$, this can be evaluated in $\mathcal{O}(N^{5})$.
    The working equations are presented in reference
    \citenum{Lesiuk2022}.
    \item Compute the SVD of $t_{ia,BC}$, and take the largest $n_{proj}$ left singular vectors as the next $V^{A}_{ia}$. This can be done in $\mathcal{O}(N^{5})$ time using a modified
    variant of truncated SVD, given in reference \citenum{Halko2011}. In this algorithm, we save
    the singular values of this step ($\sigma_{A}$), when we
    perform the CP decomposition of the triples projector. Pseudocode for this will be presented in Section IV.
    \item Iterate until convergence. Convergence is defined when
    the difference between the Frobenius norm of the rank-reduced triples amplitudes $t_{ABC}$, defined as
    \begin{equation}
        t_{ABC} = V^{A}_{ia}t_{ia,BC}
    \end{equation}
    between two successive iterations, falls below $10^{-5}$.
\end{itemize}

Since the source of the orthogonal projectors is not relevant to
the scope of this paper, we will only present results from computations utilizing the MP2 projector for the doubles amplitudes, and Lesiuk's HO-OI approach for the perturbative triples amplitudes.

\subsection{Tensor Hypercontraction (THC)}

Tensor hypercontraction (THC) can be viewed as a ``double approximation,'' where two auxiliary indices are introduced to fit a high-dimensional
tensor instead of just one. The THC form of electron repulsion integrals is defined as \cite{Hohenstein2012}:
\begin{equation}
    (pq|rs) \approx 
    x_{p}^{I}x_{q}^{I}Z^{IJ}x_{r}^{J}x_{s}^{J}
\end{equation}
This can be derived from the CP decomposition of $B^{Q}_{ia}$, (Equation \ref{eq:df-ri-approx})
\begin{align}
    B^{Q}_{ia} \approx 
    x_{i}^{I} x_{j}^{I} \eta^{QI} \\
    Z^{IJ} = 
    \eta^{QI}\eta^{QJ} \ .
\end{align}
Similarly, the THC form of coupled-cluster amplitudes can be derived from
the tensor hypercontraction of the orthogonal projectors, given by, in the case of the doubles projector:\cite{Hohenstein2022}
\begin{equation}
    U^{V}_{ia} = 
    y_{i}^{X}y_{a}^{X}\tau_{VX} \label{eq:33} \ .
\end{equation}

For the triples projector, it assumes a very similar form,

\begin{equation}
    V^{A}_{ia} = z_{i}^{X}z_{a}^{X}\theta_{AX}
\end{equation}

A PARAFAC/CANDENCOMP (CP) decomposition approach on $V^{A}_{ia}$ may be used. This approach is not dependent on the source of
the projectors, and any of the projector building approaches from
Section C may be used. Here we use the variant of CP
decomposition, first introduced by Hohenstein et. al. for the
doubles projector \cite{Hohenstein2022}, where the eigenvalues of
the doubles projector are in the CP decomposition, into the
alternating least-squares (ALS) iterations.

In our algorithm, for the decomposition of the triples amplitude,
instead of using the eigenvalues of the doubles projector, we use
the singular values of the $t_{ia,BC}$ intermediate ($\sigma_{A}$).
The functional to minimize is hence:

\begin{equation}
    L_{CP} = \sum_{ia} (\sigma_{A}[V^{A}_{ia} - z_{i}^{X}z_{a}^{X}\theta_{AX}])^{2}
\end{equation}

And the update rule for each intermediate is given as

\begin{align}
    z_{i}^{X} &= 
    \sum_{aA} \sigma_{A}^{2}V^{A}_{ia} \sum_{Y} z_{a}^{Y}\theta_{AY}
    [\sum_{b}z_{b}^{X}z_{b}^{Y}\sum_{B} \sigma_{B}^{2}\theta_{BX}\theta_{BY}]^{-1}
    \label{eq:update_zix} \\
    z_{a}^{X} &= 
    \sum_{iA} \sigma_{A}^{2}V^{A}_{ia} \sum_{Y} z_{i}^{Y}\theta_{AY}
    [\sum_{j}z_{j}^{X}z_{j}^{Y}\sum_{B} \sigma_{B}^{2}\theta_{BX}\theta_{BY}]^{-1} \label{eq:update_zax} \\
    \theta_{AX} &= 
    \sum_{ia} V^{A}_{ia} \sum_{Y} z_{i}^{Y}z_{a}^{Y}
    [\sum_{j}z_{j}^{X}z_{j}^{Y}\sum_{b}z_{b}^{X}z_{b}^{Y}]^{-1}
    \label{eq:update_theta}
\end{align}

Note that the update rule for $\theta$ is the same as in traditional
CP decomposition. \\

Since a CP decomposition does not exactly
recreate the original projector, the projectors lose their orthogonal
property \cite{Hohenstein2022}. Therefore, we have to re-create the
projectors after the CP decomposition:
\begin{align}
    S_{AB} &= 
    V^{A}_{ia} V^{B}_{ia} \\
    \theta_{AX} &= 
    \theta_{BX} S_{AB}^{-1/2} \\
    V^{A}_{ia} &= 
    z_{i}^{X} z_{a}^{X} \theta_{AX}
\end{align}

The $t_{ijk}^{abc}$ amplitudes can now be rewritten as, from equation \ref{eq:8}:

\begin{align}
    t_{ijk}^{abc} &= 
    z_{i}^{X}z_{a}^{X}z_{j}^{Y}z_{b}^{Y}z_{k}^{Z}z_{c}^{Z}t_{XYZ} 
    \label{eq:thc_triples} \\
    t_{XYZ} &= \theta_{AX}\theta_{BY}\theta_{CZ}t_{ABC}
\end{align}

Recently, Hohenstein et. al. have devised an algorithm that takes advantage
of the THC form of the $t_{ij}^{ab}$ amplitudes to develop an $O(N^{4})$ scaling
implementation of CCSD \cite{Hohenstein2022}. In the next section, we will show how to extend this to the (T) correction with the THC form
of the $t_{ijk}^{abc}$ amplitudes.

\section{Derivation of Working Equations}

We first define a couple of intermediates. From Lesiuk
\cite{Lesiuk2022}, we define:

\begin{equation}\label{eq:Dint}
    D^{QV}_{jb} = (B^{Q}_{bd}U^{W}_{jd} - B^{Q}_{lj}U^{W}_{lb})T^{VW}
\end{equation}

Next, we define the following chain of intermediates from contracting
the polyadic vectors ($z_{i}^{X}$ and $z_{a}^{X}$) of the triples projector
with the the doubles projector, the DF/RI or CD decomposed ERIs, the D
intermediate from equation \ref{eq:Dint}, as well as the $T_{1}$
amplitudes.

\begin{align}
    \overline{U}^{VX} &= U^{V}_{ia}z_{i}^{X}z_{a}^{X} \\
    \widetilde{U}^{VXY} &= U^{V}_{ia}z_{i}^{X}z_{a}^{Y} \\
    \overline{B}^{QX} &= B^{Q}_{ia}z_{i}^{X}z_{a}^{X} \\
    \widetilde{B}^{QXY} &= B^{Q}_{ia}z_{i}^{X}z_{a}^{Y} \\
    \overline{D}^{QVX} &= D^{QV}_{ia}z_{i}^{X}z_{a}^{X} \\
    \widetilde{D}^{QVXY} &= D^{QV}_{ia}z_{i}^{X}z_{a}^{Y} \\
    \overline{t_{1}}^{X} &= t_{i}^{a}z_{i}^{X}z_{a}^{X} \\
    \widetilde{t_{1}}^{XY} &= t_{i}^{a}z_{i}^{X}z_{a}^{Y}
\end{align}

We then take Equation \ref{eq:pTenergynew}, Equation \ref{eq:Vijkabc},
Equation \ref{eq:thc_triples}, and the previously defined
intermediates, to arrive at the THC form of the
triples energy correction:

\begin{align}
    E^{(T)} \mathrel{+}= 8 \cdot \overline{U}^{VX}\overline{D}^{VQY}\overline{B}^{QZ}t_{XYZ}
    \label{eq:56} \\
    E^{(T)} \mathrel{+}= 4 \cdot \overline{t_{1}}^{X}\overline{B}^{QY}\overline{B}^{QZ}t_{XYZ}
    \label{eq:57} \\
    E^{(T)} \mathrel{-}= 4 \cdot \widetilde{U}^{VXZ}\overline{D}^{VQY}\widetilde{B}^{QZX}t_{XYZ}
    \label{eq:58} \\
    E^{(T)} \mathrel{-}= 4 \cdot \widetilde{U}^{VXZ}\widetilde{D}^{VQZX}\overline{B}^{QY}t_{XYZ}
    \label{eq:59} \\
    E^{(T)} \mathrel{-}= 4 \cdot \overline{U}^{VX}\widetilde{D}^{VQXZ}\widetilde{B}^{QZX}t_{XYZ}
    \label{eq:60} \\
    E^{(T)} \mathrel{-}= 4 \cdot \widetilde{t_{1}}^{XZ}\widetilde{B}^{QZX}\overline{B}^{QY}t_{XYZ}
    \label{eq:61} \\
    E^{(T)} \mathrel{-}= 2 \cdot \overline{t_{1}}^{Y}\widetilde{B}^{QXZ}\widetilde{B}^{QZX}t_{XYZ}
    \label{eq:62} \\
    E^{(T)} \mathrel{+}= 2 \cdot \widetilde{U}^{VXZ}\widetilde{D}^{VQYX}\widetilde{B}^{QZY}t_{XYZ}
    \label{eq:63} \\
    E^{(T)} \mathrel{+}= 2 \cdot \widetilde{U}^{VXZ}\widetilde{D}^{VQZY}\widetilde{B}^{QYX}t_{XYZ}
    \label{eq:64} \\
    E^{(T)} \mathrel{+}= 2 \cdot \widetilde{t_{1}}^{XZ}\widetilde{B}^{QYX}\widetilde{B}^{QZY}t_{XYZ}
    \label{eq:65}
\end{align}

Equations \ref{eq:56} and \ref{eq:57} correspond to the first term in
equation \ref{eq:pTenergynew}, equations \ref{eq:58} through \ref{eq:62} the
second term, and equations \ref{eq:63} to \ref{eq:65} the third term. All
of the contractions can be determined in $\mathcal{O}(N^{5})$ time or less.

\section{Implementation Details}

To aid future research and development, we present pseudocode for
some of the algorithms we use for the optimal contraction of
intermediate terms to evaluate the THC-CCSD(T) energy. We first
present our non-iterative SVD algorithm to factorize
the $t_{ia,BC}$ intermediate, inspired by the truncated SVD and
diagonalization algorithms given in Ref. \citenum{Halko2011}. In Algorithm \ref{alg:t-svd}, we present a non-iterative truncated SVD algorithm
to avoid the $O(N^{6})$ scaling of a traditional SVD of the $t_{ia,BC}$ intermediate. In Algorithms \ref{alg:thc_pT_algo_1}-\ref{alg:thc_pT_algo_3}, we present suggested
contraction orders, as well as tensor slicings, for each term of
the THC-CCSD(T) energy expression. We try to make the
contractions such that highly-efficient level 3 BLAS
matrix multiplication calls are utilized as much as possible.
For each step of each algorithm, the runtime is given, and if a
level 3 BLAS matrix multiplication call is possible, then the term
(GEMM) is added. Additionally, the $\widetilde{D}^{QVXY}$ intermediate is
never fully built to help with memory costs. The runtime of this algorithm
is $\mathcal{O}(N^{5})$, with $\mathcal{O}(N^{4})$ storage costs, 
the only quartic memory requirements involve the storage of the
$t_{ia,BC}$ and  $D^{QV}_{jb}$ intermediates. It may be possible
to reduce the memory cost in future implementations of this method,
but that is beyond the scope of this paper.

\begin{algorithm}[H]
\caption{Truncated SVD algorithm for $t_{ia,BC}$}\label{alg:t-svd}
\begin{algorithmic}
\State $\Omega_{BC,X} = \textbf{random}(n_{proj} * n_{proj}, n_{proj})$
\Comment{$\mathcal{O}({N^{3}})$}
\State $Y_{ia,X} = t_{ia,BC}\Omega_{BC, X}$
\Comment{$\mathcal{O}(N^{5})$, GEMM}
\State $Q_{ia,X}, R_{X, Y} = \textbf{QR}(Y_{ia,X})$
\Comment{$\mathcal{O}(N^{4})$}
\State $t'_{X,BC} = Q_{ia,X} t_{ia,BC}$
\Comment{$\mathcal{O}(N^{5})$, GEMM}
\State $X_{XY} = t'_{X,BC}t'_{Y,BC}$
\Comment{$\mathcal{O}(N^{4})$, GEMM}
\State $V'_{XY}, \epsilon_{Y} = \textbf{diagonalize}(X_{XY})$
\Comment{$\mathcal{O}(N^{3})$}
\State $V^{A}_{ia} = Q_{ia,B} V'_{BA}$
\Comment{$\mathcal{O}(N^{4})$, GEMM}
\State $\sigma_{A} = \sqrt{\epsilon_{A}}$
\Comment{$\mathcal{O}(N)$}
\State $\textbf{return } V^{A}_{ia}, \sigma_{A}$

\end{algorithmic}
\end{algorithm}

\begin{algorithm}[H]
\caption{$E_{1}^{(T)}$ Contractions
(Equations \ref{eq:56} - \ref{eq:57})}\label{alg:thc_pT_algo_1}
\begin{algorithmic}
\State $A_{VYZ} = \overline{D}^{QVY}\overline{B}^{QZ}$
\Comment{$\mathcal{O}(N^{4})$, GEMM}
\State $B_{XYZ} = \overline{U}^{VX}\overline{A}^{VYZ}$
\Comment{$\mathcal{O}(N^{4})$, GEMM}
\State $E^{(T)} \mathrel{+}= 8 \cdot B_{XYZ} t_{XYZ}$
\Comment{$\mathcal{O}(N^{3})$}
\State $C_{YZ} = \overline{B}^{QY}\overline{B}^{QZ}$
\Comment{$\mathcal{O}(N^{3})$, GEMM}
\State $D_{X} = C_{YZ}t_{XYZ}$
\Comment{$\mathcal{O}(N^{3})$}
\State $E^{(T)} \mathrel{+}= 4 \cdot D_{X} \overline{t_{1}}^{X}$
\Comment{$\mathcal{O}(N)$}

\end{algorithmic}
\end{algorithm}

\begin{algorithm}[H]
\caption{$E_{2}^{(T)}$ Contractions
(Equations \ref{eq:58} - \ref{eq:62})}\label{alg:thc_pT_algo_2}
\begin{algorithmic}

\For{$V \texttt{ in } [0, n_{proj})$} \Comment{parallelize}
    \State $\widetilde{D}^{QVXY} = D^{QV}_{ia}z_{i}^{X}z_{a}^{Y}$
    \Comment{$\mathcal{O}(N^{5})$, GEMM, built on the fly to save
    storage}
    \State $A^{YZX} = \overline{D}^{QVY}\widetilde{B}^{QZX}$
    \Comment{$\mathcal{O}(N^{5})$, GEMM}
    \State $B^{YZX} = A^{YZX}\widetilde{U}^{VXZ}$
    \Comment{$\mathcal{O}(N^{4})$}
    \State $E^{(T)} \mathrel{-}= 4 \cdot B^{XYZ}t_{XYZ}$
    \Comment{$\mathcal{O}(N^{4})$}
    \State $C^{ZXY} = \widetilde{D}^{QVZX}\overline{B}^{QY}$
    \Comment{$\mathcal{O}(N^{5})$, GEMM}
    \State $D^{ZXY} = \widetilde{U}^{VXZ}C^{ZXY}$
    \Comment{$\mathcal{O}(N^{4})$}
    \State $E^{(T)} \mathrel{-}= 4 \cdot D^{XYZ}t_{XYZ}$
    \Comment{$\mathcal{O}(N^{4})$}
\EndFor

\For{$Q \texttt{ in } [0, n_{aux})$} \Comment{parallelize}
    \State $\widetilde{D}^{QVXY} = D^{QV}_{ia}z_{i}^{X}z_{a}^{Y}$
    \Comment{$\mathcal{O}(N^{5})$, GEMM, built on the fly to save
    storage}
    \State $F^{YXZ} = \overline{U}^{VY}\widetilde{D}^{QVXZ}$
    \Comment{$\mathcal{O}(N^{5})$, GEMM}
    \State $G^{YXZ} = F^{YXZ}\widetilde{B}^{QZX}$
    \Comment{$\mathcal{O}(N^{4})$}
    \State $E^{(T)} \mathrel{-}= 4 \cdot G^{XYZ}t_{XYZ}$
    \Comment{$\mathcal{O}(N^{4})$}
\EndFor

\State $H^{ZXY} = \widetilde{B}^{QZX}\overline{B}^{QY}$
\Comment{$\mathcal{O}(N^{4})$, GEMM}
\State $I^{ZXY} = H^{ZXY} \widetilde{t_{1}}^{XZ}$
\Comment{$\mathcal{O}(N^{3})$}
\State $E^{(T)} \mathrel{-}= 4 \cdot I^{XYZ}t_{XYZ}$
\Comment{$\mathcal{O}(N^{3})$}

\State $J^{XY} = t_{XYZ}\overline{t_{1}}^{Z}$
\Comment{$\mathcal{O}(N^{3})$}
\State $K^{XY} = \widetilde{B}^{QXY}\widetilde{B}^{QYX}$
\Comment{$\mathcal{O}(N^{3})$}
\State $E^{(T)} \mathrel{-}= 2 \cdot J^{XY}K^{XY}$
\Comment{$\mathcal{O}(N^{2})$}

\end{algorithmic}
\end{algorithm}

\begin{algorithm}[H]
\caption{$E_{3}^{(T)}$ Contractions
(Equations \ref{eq:63} - \ref{eq:65})}\label{alg:thc_pT_algo_3}
\begin{algorithmic}

\For{$V \texttt{ in } [0, n_{proj})$} \Comment{parallelize}
    \State $\widetilde{D}^{QVXY} = D^{QV}_{ia}z_{i}^{X}z_{a}^{Y}$
    \Comment{$\mathcal{O}(N^{5})$, GEMM, built on the fly to
    save storage}
    \For{$Y \texttt{ in } [0, n_{proj})$}
        \State $A^{XZ} = \widetilde{D}^{QVYX}\widetilde{B}^{QZY}$ \Comment{$\mathcal{O}(N^{5})$, GEMM}
        \State $B^{XZ} = A^{XZ}U^{VXZ}$
        \Comment{$\mathcal{O}(N^{4})$}
        \State $E^{(T)} \mathrel{+}= 2 \cdot B^{XZ}T^{XYZ}$
        \Comment{$\mathcal{O}(N^{4})$}
        
        \State $C^{XZ} = \widetilde{D}^{QVZY}\widetilde{B}^{QYX}$ \Comment{$\mathcal{O}(N^{5})$, GEMM}
        \State $D^{XZ} = C^{XZ}U^{VXZ}$
        \Comment{$\mathcal{O}(N^{4})$}
        \State $E^{(T)} \mathrel{+}= 2 \cdot D^{XZ}T^{XYZ}$
        \Comment{$\mathcal{O}(N^{4})$}
    \EndFor
\EndFor

\For{$Y \texttt{ in } [0, n_{proj})$} \Comment{parallelize}
    \State $F^{XZ} = \widetilde{B}^{QYX}\widetilde{B}^{QZY}$ \Comment{$\mathcal{O}(N^{4})$, GEMM}
    \State $G^{XZ} = F^{XZ}\widetilde{t_{1}}^{XZ}$
    \Comment{$\mathcal{O}(N^{3})$}
    \State $E^{(T)} \mathrel{+}= 2 \cdot G^{XZ}T^{XYZ}$
    \Comment{$\mathcal{O}(N^{3})$}
\EndFor

\end{algorithmic}
\end{algorithm}

The code is implemented in a developmental plugin version of the \textsc{Psi4} Quantum
Chemistry code \cite{Smith2020}, following the completion of an exact CCSD computation. Tensor contractions are performed with
the help of the EinsumsInCpp software (public on GitHub). The compressed doubles amplitudes $T^{VW}$ used to build the
triples projector are formed by transforming the exact CCSD amplitudes from
the preceding computation by the MP2 projector amplitudes.
This method is designed to be fully compatible and used with Hohenstein's
THC-CCSD method \cite{Hohenstein2022}. Future studies of using
THC-CCSD(T) in conjunction with THC-CCSD is encouraged.
\newpage

\section{Results}

\subsection{Conformation Energies}

We first evaluate our new THC-CCSD(T) method on the CYCONF
\cite{Wilke2009} \cite{Goerigk2010} data set, a set containing 11
different conformations of gaseous cysteine, with 10 corresponding
conformation energies, relative to the lowest conformer. We evaluate
conformation energies for each of the 10 conformations in CCSD, CCSD(T),
and THC-CCSD(T), and for each system, and we use the exact CCSD(T) conformation
energy as the reference. We do this using the cc-pVDZ and jun-cc-pVDZ
Dunning correlation-consistent basis sets
\cite{Dunning1989,Woon1993,Woon1994,Papajak2011}. The basis set
jun-cc-pVDZ consists of diffuse functions added to all heavy atoms,
except for the basis functions with the highest angular momentum. For the
THC-CCSD(T) computations, we set the eigenvalue
tolerance of the MP2 projector to be $10^{-4}$. In other words, the
ranks ($n_{proj}$) of the doubles and triples projectors are determined from
how many eigenvalues of the MP2 $t_{ij}^{ab}$ amplitudes are greater
than $10^{-4}$, defined as $\tau$ from Equation \ref{eq:29} in our work.
For these computations, $n_{proj}$ is around 400, compared to the max
possible rank of 2205 ($n_{occ}n_{virt}$) in the cc-pVDZ basis, yielding a
compression ratio of around 18\%. Similarly, in the jun-cc-pVDZ basis, the
ratio is 440/2793, which is around 16\%.

The summary statistics are presented in Table I, and the results for
each individual conformation are presented in Figure 1. In the table,
for the THC-CCSD(T) algorithms, the eigenvalue tolerance is given in
parentheses. To summarize the findings, THC-CCSD(T) consistently gives
lower errors compared to CCSD, for both basis sets, and the errors are
on the order of less than $0.1$ kcal/mol. It is further encouraging to
note that the absolute energy errors for these sets of computations
hover around $0.3-0.4$ kcal/mol, such that the evaluation of relative
energies benefits from favorable error cancellation. The error also does
not significantly grow with the addition of diffuse functions, from cc-pVDZ
to jun-cc-pVDZ.
\newpage

\begin{table}[H]\label{table:conf-errors}
    \caption{\label{tab:widgets}Errors in conformation energy compared to the exact CCSD(T) reference (kcal/mol). The number in parenthesis is the eigenvalue tolerance used to determine projector rank.}
    \begin{tabularx}{\textwidth}{| l | X | X | X | X |}
    \hline
    Test Set & Mean Error & MAE & RMSE & Std Dev \\\hline
    CCSD/cc-pVDZ & $-0.343$ & 0.343 & 0.384 & 0.173 \\
    THC-CCSD(T)/cc-pVDZ ($10^{-4}$) & $-0.072$ & 0.072 & 0.075 & 0.023 \\
    CCSD/jun-cc-pVDZ & $-0.291$ & 0.291 & 0.323 & 0.141 \\
    THC-CCSD(T)/jun-cc-pVDZ ($10^{-4}$) & $-0.076$ & 0.076 & 0.082 & 0.031 \\
    \hline
    \end{tabularx}
\end{table}

\begin{figure}[H]
    \centering
    \includegraphics[width=\textwidth]{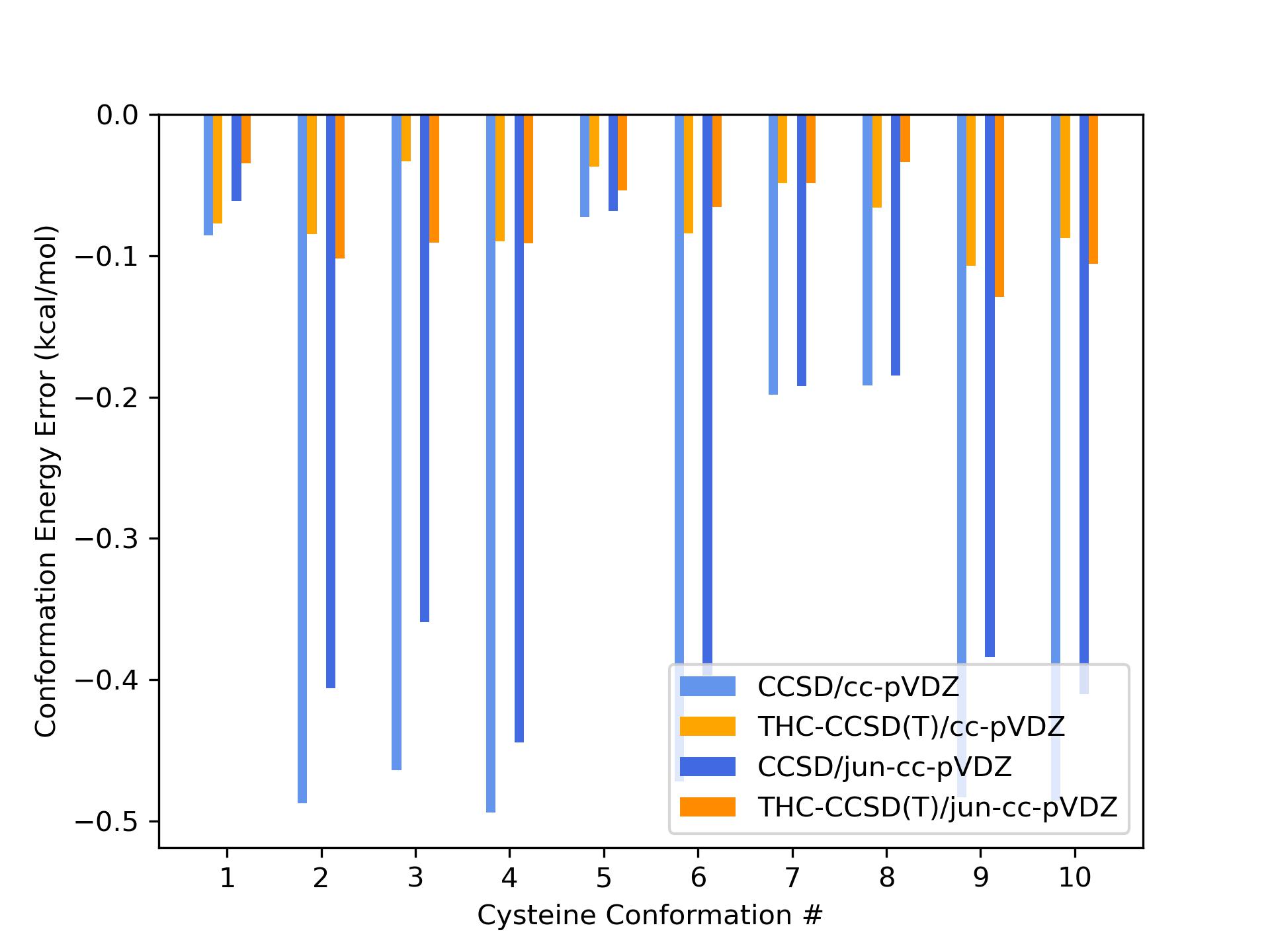}
    \caption{Errors in conformation energies for CCSD and THC-CCSD(T)
    evaluated on the CYCONF data set, compared to the exact CCSD(T)
    reference, evaluated in with the cc-pVDZ and jun-cc-pVDZ basis
    sets, with a $10^{-4}$ eigenvalue tolerance.}
    \label{fig:conf-errors}
\end{figure}

\newpage

\subsection{Potential Energy Surface}

We perform next, a potential energy surface scan on the benzene-HCN dimer
system (compound 19 from the on S22 data set \cite{Jureka2006}),
with the hydrogen atom of HCN pointing towards the $\pi$-bonds in the
benzene. We measured the energy of the system at five different
inter-atomic distances, relative to the equilibrium geometry, ranging
from 0.9 to 2.0 times the equilibrium geometry length, with the
geometries coming from the S22x5 data set \cite{Grafova2010}. In Figure
2, we plot the shape of the potential energy surface of the THC-CCSD(T)
method at an eigenvalue tolerance of $10^{-4}$, as well as using
predetermined projector ranks of 400 and 500. For all systems, an
eigenvalue tolerance of $10^{-4}$ corresponds to a projector rank
between 420-430. All THC-CCSD(T) computations better capture the
potential energy surface than the reference CCSD computations, with the
computations with the predetermined projector ranks better capturing the
shape of the surface than the one with a set eigenvalue tolerance.
The THC-CCSD(T) potential energy surface with $n_{proj}$ set to 500
exactly matches the CCSD(T) potential energy surface, for practical
purposes, with a max error of 0.027 kcal/mol, and a RMSE of 0.014 kcal/mol. The shape of the potential energy surface, for each method,
is shown in Figure 2, while the error statistics are presented in
Table II. The errors are especially
encouraging for the case of $n_{proj}$ set to 500, as the
absolute energy error of each system
compared to CCSD(T) hover around 0.4 kcal/mol.

\begin{table}[H]
    \caption{\label{tab:widgets}Errors in relative energies compared
    to the exact CCSD(T) reference (kcal/mol), for a reference CCSD
    computation, as well as THC-CCSD(T) computations with varying parameters.}
    \begin{tabularx}{\textwidth}{| l | X | X | X | X |}
    \hline
    Test Set & Mean Error & MAE & RMSE & Std Dev \\\hline
    CCSD & $-0.138$ & 0.200 & 0.236 & 0.191 \\
    THC-CCSD(T), tol = $10^{-4}$ & $-0.100$ & 0.103 & 0.132 & 0.086 \\
    THC-CCSD(T), $n_{proj} = 400$ & $-0.098$ & 0.098 & 0.128 & 0.082 \\
    THC-CCSD(T), $n_{proj} = 500$ & $-0.001$ & 0.010 & 0.014 & 0.014 \\
    \hline
    \end{tabularx}
\end{table}

\newpage

\begin{figure}[H]
    \centering
    \includegraphics[width=\textwidth]{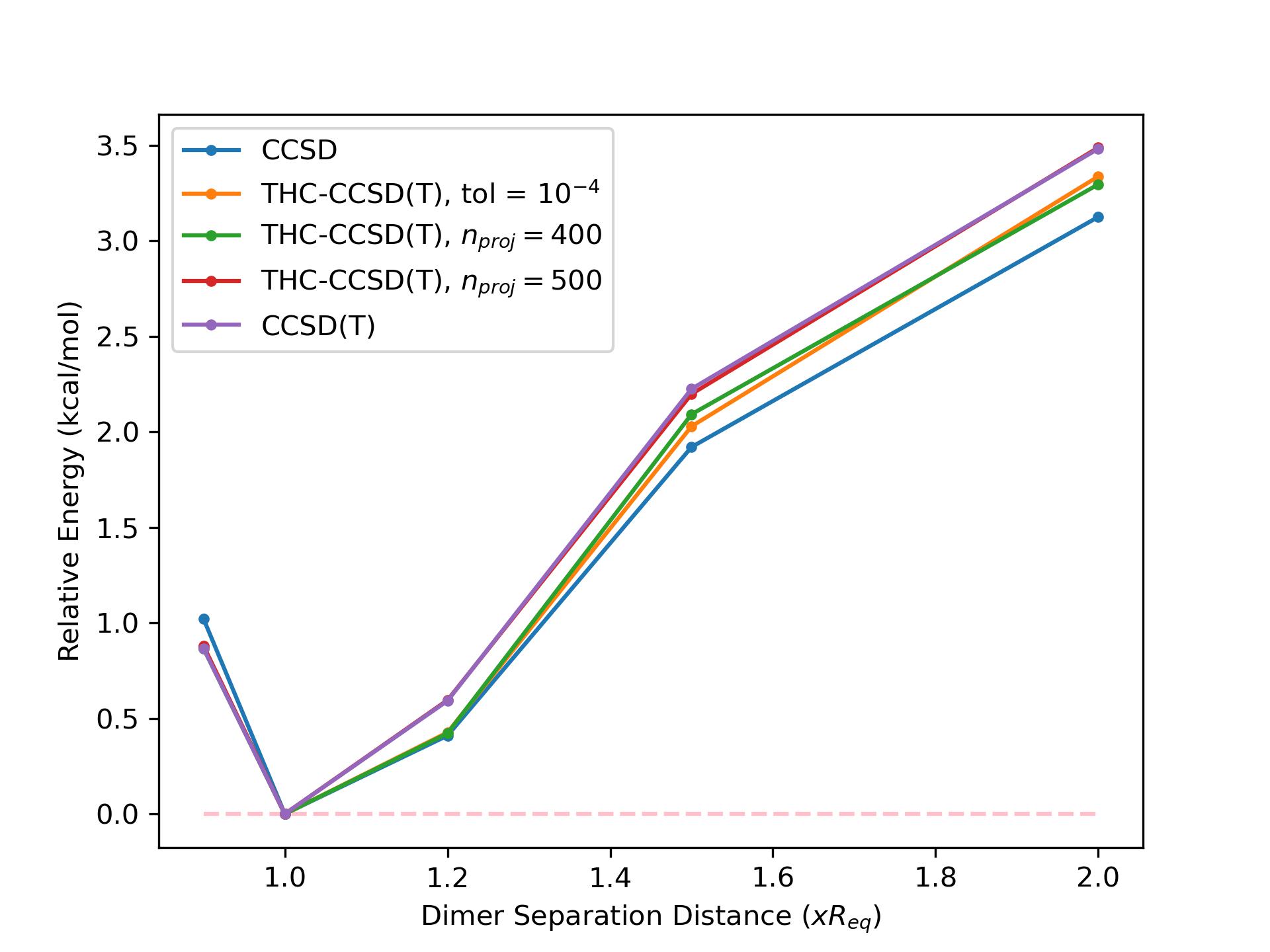}
    \caption{The relative energies of benzene-HCN dimer
    (S22 system 19) evaluated with each method at five different
    dimer separation distances relative to the equilibrium geometry.}
    \label{fig:pe-surface}
\end{figure}
\newpage

\subsection{Rank Convergence}

Next, to demonstrate the convergence of the THC-CCSD(T) method, compared
to the exact CCSD(T) energy, we ran a series of computations of the water dimer from the
S22 set \cite{Jureka2006}, at eigenvalue tolerances from $10^{-3}$ to
$10^{-11}$. An eigenvalue tolerance of $10^{-11}$ corresponds to no rank
compression for this system. The errors with respect to eigenvalue
tolerance and compression ranks are plotted in Figure 3, and it is
encouraging to see the errors decrease smoothly to the true CCSD(T)
energy, within the DF/RI approximation of the ERIs. We attribute the
``kink'' in the graph from $10^{-4}$ to $10^{-6}$ as an artifact of the CP
decomposition of the triples projector, with the CP error increasing
slightly between the projector ranks of 122 - 156, before going back
down. This artifact is well known on studies of the CP decomposition
algorithm \cite{Kolda2009}, where medium CP decomposition ranks suffer
larger losses in accuracy compared to small or large ranks. Further studies
and work are encouraged to look for ways to mitigate this phenomenon in
the context of decomposing CC amplitudes.

\begin{figure}[!]
    \centering
    \includegraphics[width=\textwidth]{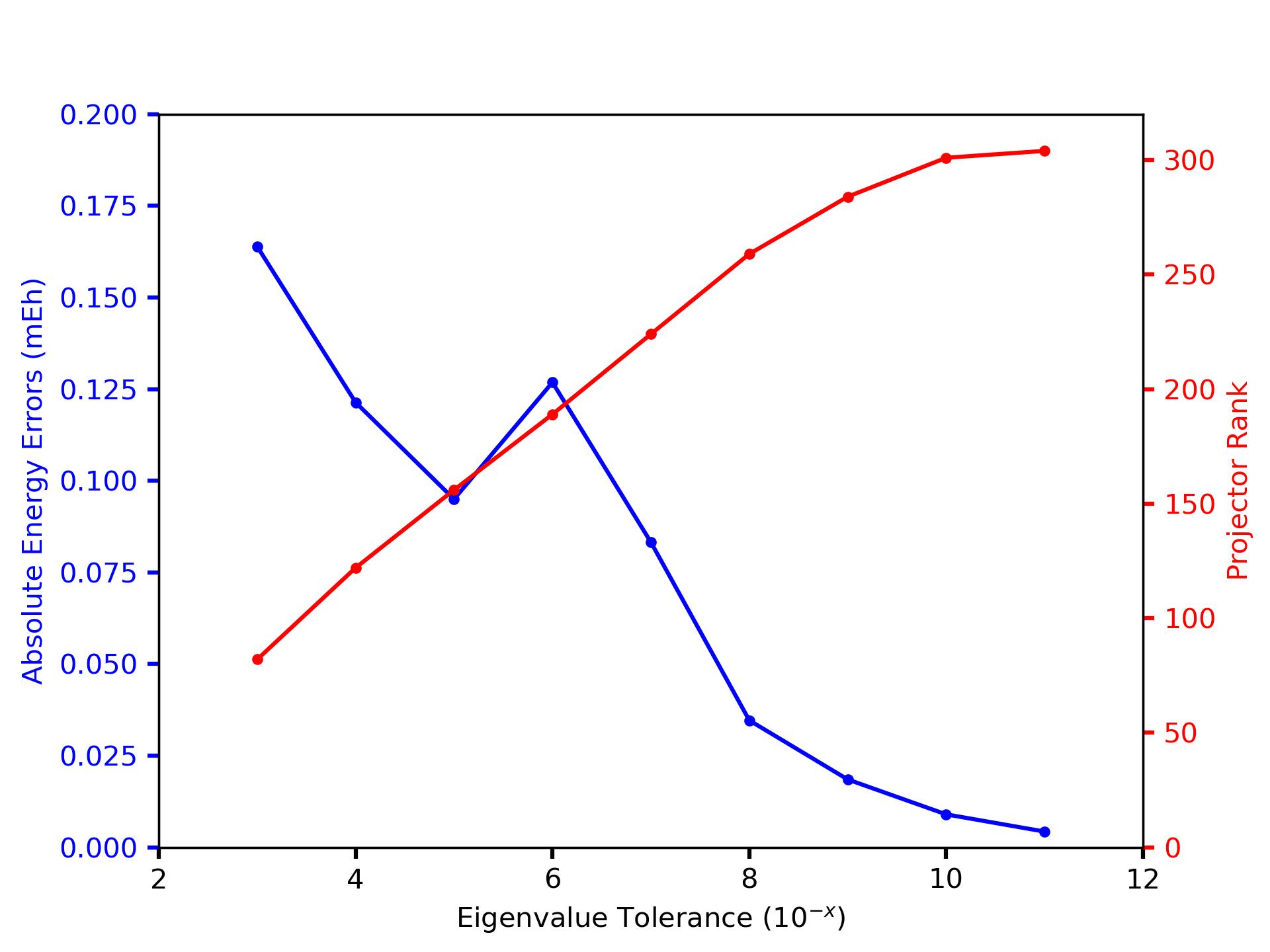}
    \caption{The convergence of the absolute energy of a water
    dimer system (S22), with respect to eigenvalue tolerance and rank.}
    \label{fig:thc-convergence}
\end{figure}
\newpage

\subsection{Scaling}

To establish the $\mathcal{O}(N^{5})$ scaling of the THC-CCSD(T) method,
it must be shown that the projector rank, or $n_{proj}$ must scale
linearly with respect to system size. Hohenstein
and Parrish have previously established the linear scaling of $n_{proj}$
for doubles amplitudes in their previous work
\cite{Parrish2019,Hohenstein2022}. However, to verify this in our
algorithm, we must show that the error does not grow more than linearly with
linear increases in system size. Below, we present THC-CCSD(T) computations
on systematically larger systems of waterclusters and linear alkanes, from 1-8
heavy atoms, in the cc-pVDZ and jun-cc-pVDZ basis sets, evaluated at an
eigenvalue tolerance of $10^{-4}$. As shown in Figures 4-7, sub-linear to linear error
growth are shown, with respect to projector rank and
system size, with virtually no loss in accuracy from cc-pVDZ
to jun-cc-pVDZ in both systems.

\begin{figure}[!]
    \centering
    \includegraphics[width=\textwidth]{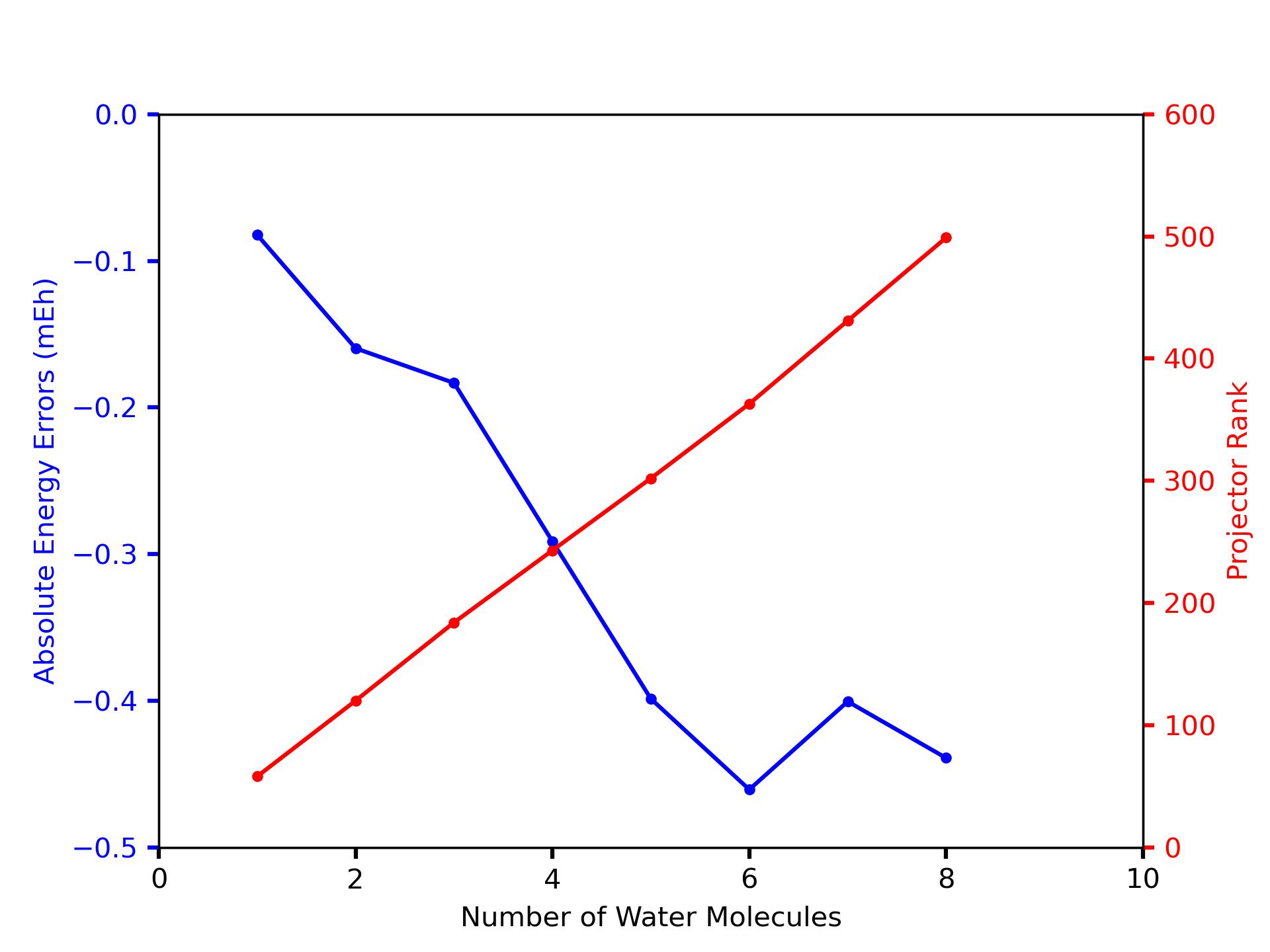}
    \caption{The growth of the absolute energy error, and
    projector rank, in a growing series of water clusters
    $(H_{2}O)_{n}$, cc-pVDZ basis.}
    \label{fig:water-dz}
\end{figure}

\begin{figure}[!]
    \centering
    \includegraphics[width=\textwidth]{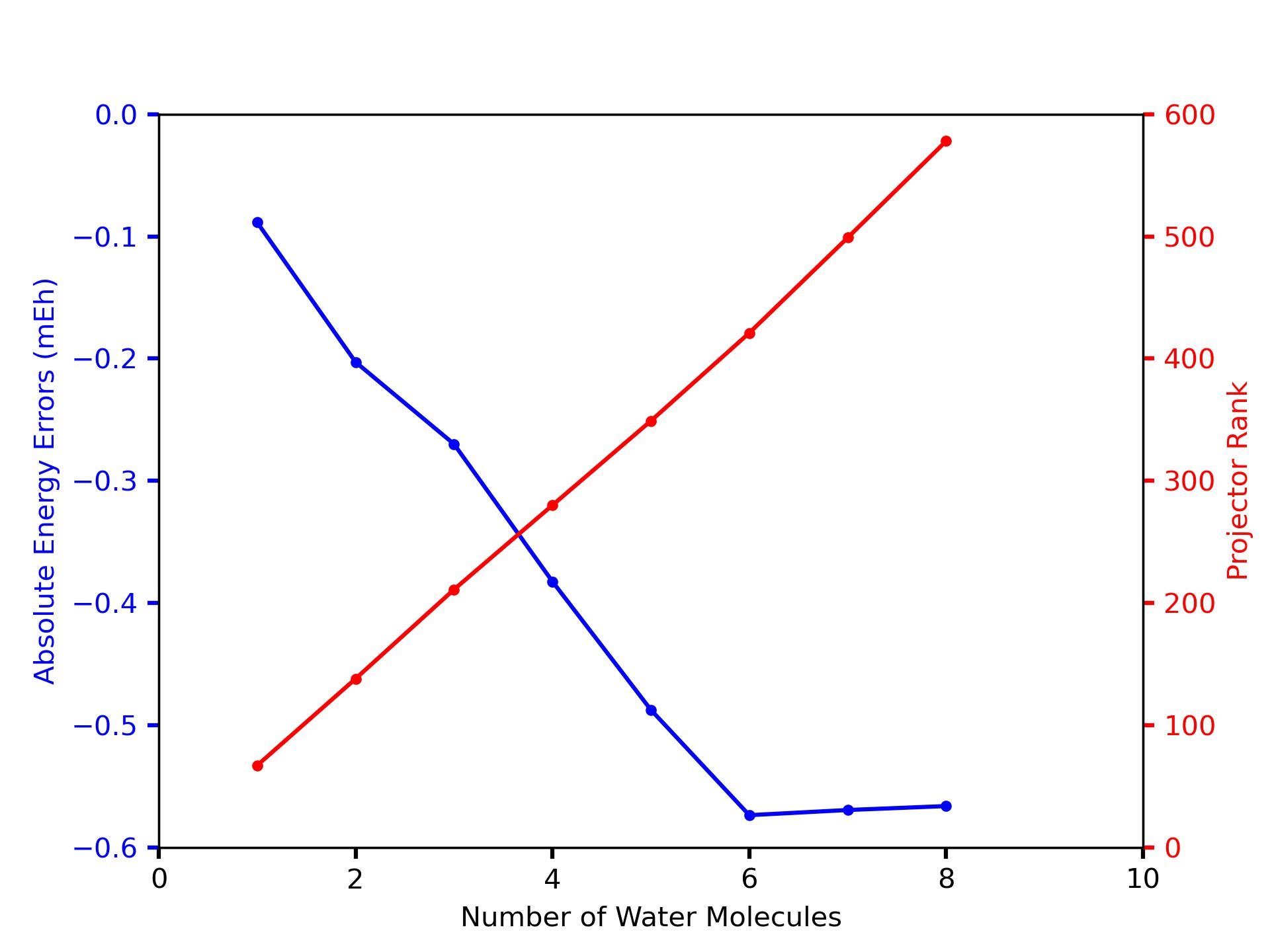}
    \caption{The growth of the absolute energy error, and
    projector rank, in a growing series of water clusters
    $(H_{2}O)_{n}$, jun-cc-pVDZ basis.}
    \label{fig:water-jdz}
\end{figure}

\begin{figure}[!]
    \centering
    \includegraphics[width=\textwidth]{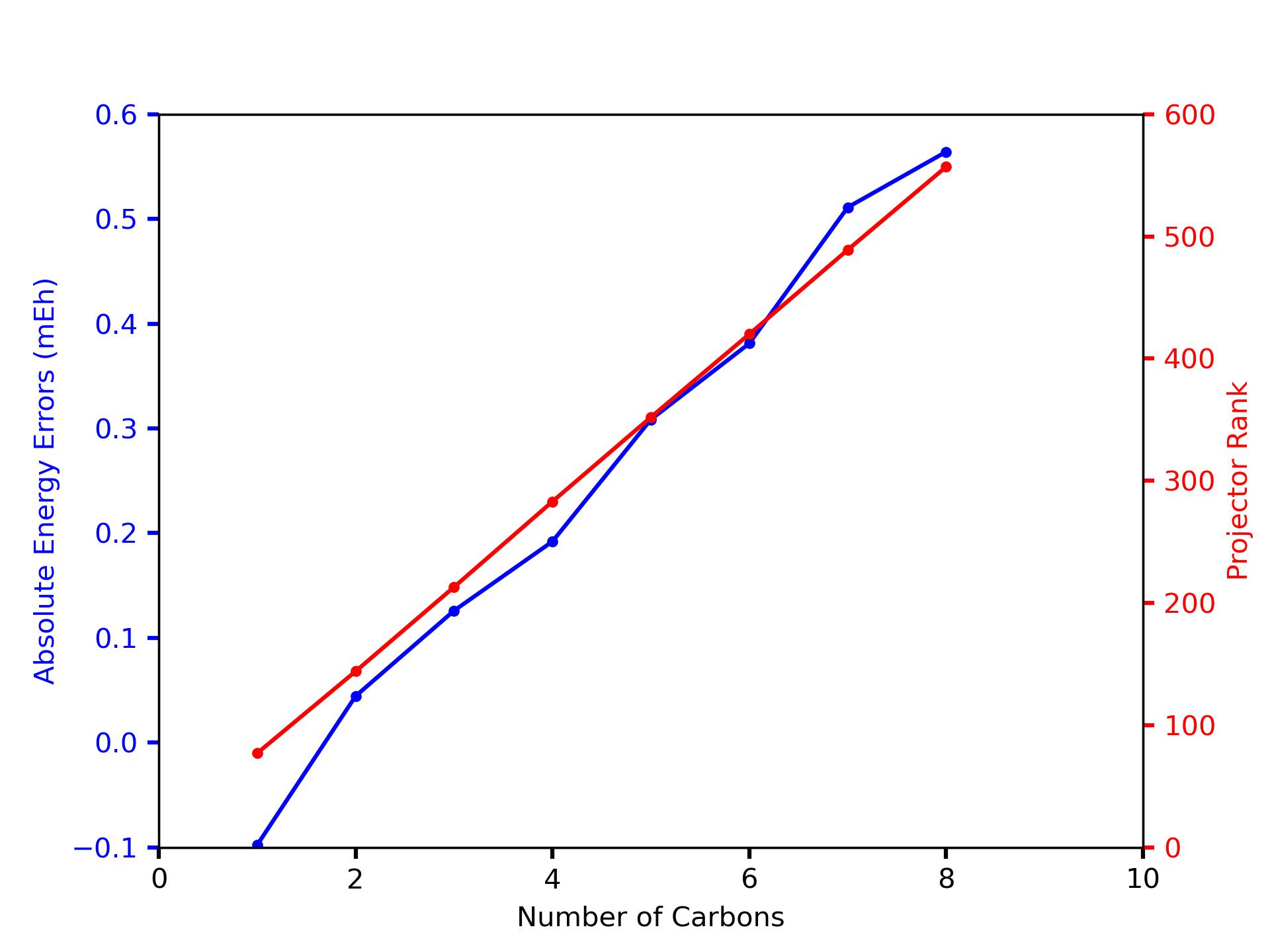}
    \caption{The growth of the absolute energy error, and
    projector rank, in a growing series of linear alkanes
    $(C_{n}H_{2n+2})$, cc-pVDZ basis.}
    \label{fig:alkane-dz}
\end{figure}

\begin{figure}[!]
    \centering
    \includegraphics[width=\textwidth]{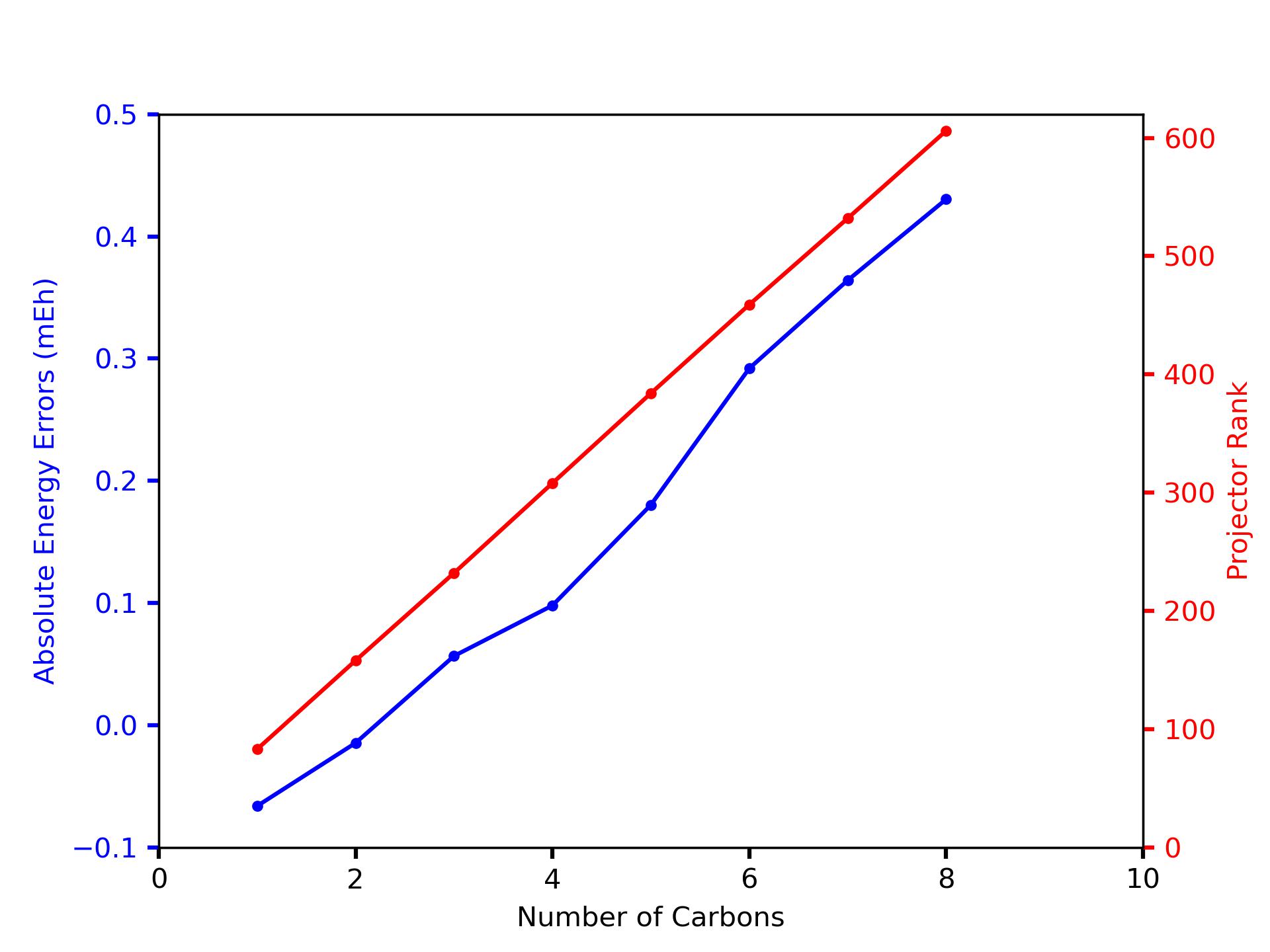}
    \caption{The growth of the absolute energy error, and
    projector rank, in a growing series of linear alkanes
    $(C_{n}H_{2n+2})$, jun-cc-pVDZ basis.}
    \label{fig:alkane-jdz}
\end{figure}
\clearpage

\section{Conclusions}

In this paper, we present the working equations for the
THC-CCSD(T) method, a $\mathcal{O}(N^{5})$ scaling approximation
to CCSD(T), that allows for systematic control of errors. In our
pilot implementation, we show the errors are controllable to the
point of maintaining chemical accuracy of less than 0.1 kcal/mol
for relative energies, and 1 mEh for absolute energies, while
maintaining size extensivity. We also showed that the
method yields continuous potential energy surfaces that closely matches
the CCSD(T) surfaces with sufficient projector rank. In the
future, we hope to consider ways to improve the errors of the method at
a given eigenvalue tolerance, such as through using other sources
for the orthogonal projector. We would also like to look into
alternative approaches to the THC factorization of orthogonal projectors. Though a CP decomposition is generally applicable,
and relatively easy to implement, it does not assume any
underlying form about the amplitudes. One avenue is the extension
of the quadrature-based approach of Parrish, Hohenstein,
Martinez, and Sherrill with Least-Squares Tensor Hypercontraction
(LS-THC) to the triples amplitudes \cite{Parrish2012,Hohenstein2012b}.

\section*{Acknowledgements}

The authors gratefully acknowledge financial support from the
U.S. Department of Energy, Basic Energy Sciences Division,
Computational and Theoretical Chemistry (CTC) Grant DE-SC0018164.

\section*{Data Availability}
The data that supports the findings of this study are available
with the article and its supplementary material.

\bibliography{citations}

\begin{thebibliography}{79}%
\makeatletter
\providecommand \@ifxundefined [1]{%
 \@ifx{#1\undefined}
}%
\providecommand \@ifnum [1]{%
 \ifnum #1\expandafter \@firstoftwo
 \else \expandafter \@secondoftwo
 \fi
}%
\providecommand \@ifx [1]{%
 \ifx #1\expandafter \@firstoftwo
 \else \expandafter \@secondoftwo
 \fi
}%
\providecommand \natexlab [1]{#1}%
\providecommand \enquote  [1]{``#1''}%
\providecommand \bibnamefont  [1]{#1}%
\providecommand \bibfnamefont [1]{#1}%
\providecommand \citenamefont [1]{#1}%
\providecommand \href@noop [0]{\@secondoftwo}%
\providecommand \href [0]{\begingroup \@sanitize@url \@href}%
\providecommand \@href[1]{\@@startlink{#1}\@@href}%
\providecommand \@@href[1]{\endgroup#1\@@endlink}%
\providecommand \@sanitize@url [0]{\catcode `\\12\catcode `\$12\catcode
  `\&12\catcode `\#12\catcode `\^12\catcode `\_12\catcode `\%12\relax}%
\providecommand \@@startlink[1]{}%
\providecommand \@@endlink[0]{}%
\providecommand \url  [0]{\begingroup\@sanitize@url \@url }%
\providecommand \@url [1]{\endgroup\@href {#1}{\urlprefix }}%
\providecommand \urlprefix  [0]{URL }%
\providecommand \Eprint [0]{\href }%
\providecommand \doibase [0]{http://dx.doi.org/}%
\providecommand \selectlanguage [0]{\@gobble}%
\providecommand \bibinfo  [0]{\@secondoftwo}%
\providecommand \bibfield  [0]{\@secondoftwo}%
\providecommand \translation [1]{[#1]}%
\providecommand \BibitemOpen [0]{}%
\providecommand \bibitemStop [0]{}%
\providecommand \bibitemNoStop [0]{.\EOS\space}%
\providecommand \EOS [0]{\spacefactor3000\relax}%
\providecommand \BibitemShut  [1]{\csname bibitem#1\endcsname}%
\let\auto@bib@innerbib\@empty
\bibitem [{\citenamefont {Crawford}\ and\ \citenamefont {{H. F.
  Schaefer}}(2007)}]{Crawford2007}%
  \BibitemOpen
  \bibfield  {author} {\bibinfo {author} {\bibfnamefont {T.~D.}\ \bibnamefont
  {Crawford}}\ and\ \bibinfo {author} {\bibnamefont {{H. F. Schaefer}}},\
  }\href@noop {} {\bibfield  {journal} {\bibinfo  {journal} {Reviews in
  Computational Chemistry}\ ,\ \bibinfo {pages} {33}} (\bibinfo {year}
  {2007})}\BibitemShut {NoStop}%
\bibitem [{\citenamefont {Bartlett}\ and\ \citenamefont
  {Musial}(2007)}]{Bartlett2007}%
  \BibitemOpen
  \bibfield  {author} {\bibinfo {author} {\bibfnamefont {R.~J.}\ \bibnamefont
  {Bartlett}}\ and\ \bibinfo {author} {\bibfnamefont {M.}~\bibnamefont
  {Musial}},\ }\href@noop {} {\bibfield  {journal} {\bibinfo  {journal}
  {Reviews of Modern Physics}\ }\textbf {\bibinfo {volume} {79}} (\bibinfo
  {year} {2007})}\BibitemShut {NoStop}%
\bibitem [{\citenamefont {Cramer}(2002)}]{Cramer2002}%
  \BibitemOpen
  \bibfield  {author} {\bibinfo {author} {\bibfnamefont {C.~J.}\ \bibnamefont
  {Cramer}},\ }\href@noop {} {\emph {\bibinfo {title} {Essentials of
  Computational Chemistry}}}\ (\bibinfo {year} {2002})\ pp.\ \bibinfo {pages}
  {191--232}\BibitemShut {NoStop}%
\bibitem [{\citenamefont {Sherrill}\ and\ \citenamefont {{H. F.
  Schaefer}}(1999)}]{Sherrill1999}%
  \BibitemOpen
  \bibfield  {author} {\bibinfo {author} {\bibfnamefont {C.~D.}\ \bibnamefont
  {Sherrill}}\ and\ \bibinfo {author} {\bibnamefont {{H. F. Schaefer}}},\
  }\href@noop {} {\bibfield  {journal} {\bibinfo  {journal} {Advances in
  Quantum Chemistry}\ }\textbf {\bibinfo {volume} {34}},\ \bibinfo {pages}
  {143} (\bibinfo {year} {1999})}\BibitemShut {NoStop}%
\bibitem [{\citenamefont {Raghavachari}\ \emph {et~al.}(1989)\citenamefont
  {Raghavachari}, \citenamefont {Trucks}, \citenamefont {Pople},\ and\
  \citenamefont {Head-Gordon}}]{Raghavachari1989}%
  \BibitemOpen
  \bibfield  {author} {\bibinfo {author} {\bibfnamefont {K.}~\bibnamefont
  {Raghavachari}}, \bibinfo {author} {\bibfnamefont {G.~W.}\ \bibnamefont
  {Trucks}}, \bibinfo {author} {\bibfnamefont {J.~A.}\ \bibnamefont {Pople}}, \
  and\ \bibinfo {author} {\bibfnamefont {M.}~\bibnamefont {Head-Gordon}},\
  }\href {\doibase https://doi.org/10.1016/S0009-2614(89)87395-6} {\bibfield
  {journal} {\bibinfo  {journal} {Chemical Physics Letters}\ }\textbf {\bibinfo
  {volume} {157}},\ \bibinfo {pages} {479} (\bibinfo {year}
  {1989})}\BibitemShut {NoStop}%
\bibitem [{\citenamefont {Hirata}(2003)}]{Hirata2003}%
  \BibitemOpen
  \bibfield  {author} {\bibinfo {author} {\bibfnamefont {S.}~\bibnamefont
  {Hirata}},\ }\href {\doibase 10.1021/jp034596z} {\bibfield  {journal}
  {\bibinfo  {journal} {The Journal of Physical Chemistry A}\ }\textbf
  {\bibinfo {volume} {107}},\ \bibinfo {pages} {9887} (\bibinfo {year}
  {2003})}\BibitemShut {NoStop}%
\bibitem [{\citenamefont {Auer}\ \emph {et~al.}(2006)\citenamefont {Auer},
  \citenamefont {Baumgartner}, \citenamefont {Bernholdt}, \citenamefont
  {Bibireata}, \citenamefont {Choppella}, \citenamefont {Cociorva},
  \citenamefont {Gao}, \citenamefont {Harrison}, \citenamefont
  {Krishnamoorthy}, \citenamefont {Krishnan}, \citenamefont {Lam},
  \citenamefont {Lu}, \citenamefont {Nooijen}, \citenamefont {Pitzer},
  \citenamefont {Ramanujam}, \citenamefont {Sadayappan},\ and\ \citenamefont
  {Sibiryakov}}]{Auer2006}%
  \BibitemOpen
  \bibfield  {author} {\bibinfo {author} {\bibfnamefont {A.~A.}\ \bibnamefont
  {Auer}}, \bibinfo {author} {\bibfnamefont {G.}~\bibnamefont {Baumgartner}},
  \bibinfo {author} {\bibfnamefont {D.~E.}\ \bibnamefont {Bernholdt}}, \bibinfo
  {author} {\bibfnamefont {A.}~\bibnamefont {Bibireata}}, \bibinfo {author}
  {\bibfnamefont {V.}~\bibnamefont {Choppella}}, \bibinfo {author}
  {\bibfnamefont {D.}~\bibnamefont {Cociorva}}, \bibinfo {author}
  {\bibfnamefont {X.}~\bibnamefont {Gao}}, \bibinfo {author} {\bibfnamefont
  {R.}~\bibnamefont {Harrison}}, \bibinfo {author} {\bibfnamefont
  {S.}~\bibnamefont {Krishnamoorthy}}, \bibinfo {author} {\bibfnamefont
  {S.}~\bibnamefont {Krishnan}}, \bibinfo {author} {\bibfnamefont {C.-C.}\
  \bibnamefont {Lam}}, \bibinfo {author} {\bibfnamefont {Q.}~\bibnamefont
  {Lu}}, \bibinfo {author} {\bibfnamefont {M.}~\bibnamefont {Nooijen}},
  \bibinfo {author} {\bibfnamefont {R.}~\bibnamefont {Pitzer}}, \bibinfo
  {author} {\bibfnamefont {J.}~\bibnamefont {Ramanujam}}, \bibinfo {author}
  {\bibfnamefont {P.}~\bibnamefont {Sadayappan}}, \ and\ \bibinfo {author}
  {\bibfnamefont {A.}~\bibnamefont {Sibiryakov}},\ }\href {\doibase
  10.1080/00268970500275780} {\bibfield  {journal} {\bibinfo  {journal}
  {Molecular Physics}\ }\textbf {\bibinfo {volume} {104}},\ \bibinfo {pages}
  {211} (\bibinfo {year} {2006})}\BibitemShut {NoStop}%
\bibitem [{\citenamefont {Janowski}\ \emph {et~al.}(2007)\citenamefont
  {Janowski}, \citenamefont {Ford},\ and\ \citenamefont
  {Pulay}}]{Janowski2007}%
  \BibitemOpen
  \bibfield  {author} {\bibinfo {author} {\bibfnamefont {T.}~\bibnamefont
  {Janowski}}, \bibinfo {author} {\bibfnamefont {A.~R.}\ \bibnamefont {Ford}},
  \ and\ \bibinfo {author} {\bibfnamefont {P.}~\bibnamefont {Pulay}},\ }\href
  {\doibase 10.1021/ct700048u} {\bibfield  {journal} {\bibinfo  {journal}
  {Journal of Chemical Theory and Computation}\ }\textbf {\bibinfo {volume}
  {3}},\ \bibinfo {pages} {1368} (\bibinfo {year} {2007})}\BibitemShut
  {NoStop}%
\bibitem [{\citenamefont {Janowski}\ and\ \citenamefont
  {Pulay}(2008)}]{Janowski2008}%
  \BibitemOpen
  \bibfield  {author} {\bibinfo {author} {\bibfnamefont {T.}~\bibnamefont
  {Janowski}}\ and\ \bibinfo {author} {\bibfnamefont {P.}~\bibnamefont
  {Pulay}},\ }\href {\doibase 10.1021/ct800142f} {\bibfield  {journal}
  {\bibinfo  {journal} {Journal of Chemical Theory and Computation}\ }\textbf
  {\bibinfo {volume} {4}},\ \bibinfo {pages} {1585} (\bibinfo {year}
  {2008})}\BibitemShut {NoStop}%
\bibitem [{\citenamefont {van Dam}\ \emph {et~al.}(2011)\citenamefont {van
  Dam}, \citenamefont {de~Jong}, \citenamefont {Bylaska}, \citenamefont
  {Govind}, \citenamefont {Kowalski}, \citenamefont {Straatsma},\ and\
  \citenamefont {Valiev}}]{vanDam2011}%
  \BibitemOpen
  \bibfield  {author} {\bibinfo {author} {\bibfnamefont {H.}~\bibnamefont {van
  Dam}}, \bibinfo {author} {\bibfnamefont {W.}~\bibnamefont {de~Jong}},
  \bibinfo {author} {\bibfnamefont {E.}~\bibnamefont {Bylaska}}, \bibinfo
  {author} {\bibfnamefont {N.}~\bibnamefont {Govind}}, \bibinfo {author}
  {\bibfnamefont {K.}~\bibnamefont {Kowalski}}, \bibinfo {author}
  {\bibfnamefont {T.}~\bibnamefont {Straatsma}}, \ and\ \bibinfo {author}
  {\bibfnamefont {M.}~\bibnamefont {Valiev}},\ }\href {\doibase
  10.1002/wcms.62} {\bibfield  {journal} {\bibinfo  {journal} {Wiley
  Interdisciplinary Reviews: Computational Molecular Science}\ }\textbf
  {\bibinfo {volume} {1}},\ \bibinfo {pages} {888} (\bibinfo {year}
  {2011})}\BibitemShut {NoStop}%
\bibitem [{\citenamefont {Deumens}\ \emph {et~al.}(2011)\citenamefont
  {Deumens}, \citenamefont {Lotrich}, \citenamefont {Perera}, \citenamefont
  {Ponton}, \citenamefont {Sanders},\ and\ \citenamefont
  {Bartlett}}]{Deumens2011}%
  \BibitemOpen
  \bibfield  {author} {\bibinfo {author} {\bibfnamefont {E.}~\bibnamefont
  {Deumens}}, \bibinfo {author} {\bibfnamefont {V.~F.}\ \bibnamefont
  {Lotrich}}, \bibinfo {author} {\bibfnamefont {A.}~\bibnamefont {Perera}},
  \bibinfo {author} {\bibfnamefont {M.~J.}\ \bibnamefont {Ponton}}, \bibinfo
  {author} {\bibfnamefont {B.~A.}\ \bibnamefont {Sanders}}, \ and\ \bibinfo
  {author} {\bibfnamefont {R.~J.}\ \bibnamefont {Bartlett}},\ }\href {\doibase
  10.1002/wcms.77} {\bibfield  {journal} {\bibinfo  {journal} {Wiley
  Interdisciplinary Reviews: Computational Molecular Science}\ }\textbf
  {\bibinfo {volume} {1}},\ \bibinfo {pages} {895} (\bibinfo {year}
  {2011})}\BibitemShut {NoStop}%
\bibitem [{\citenamefont {Kobayashi}\ and\ \citenamefont
  {Rendell}(1997)}]{Kobayashi1997}%
  \BibitemOpen
  \bibfield  {author} {\bibinfo {author} {\bibfnamefont {R.}~\bibnamefont
  {Kobayashi}}\ and\ \bibinfo {author} {\bibfnamefont {A.~P.}\ \bibnamefont
  {Rendell}},\ }\href {\doibase 10.1016/S0009-2614(96)01387-5} {\bibfield
  {journal} {\bibinfo  {journal} {Chemical Physics Letters}\ }\textbf {\bibinfo
  {volume} {265}},\ \bibinfo {pages} {1} (\bibinfo {year} {1997})}\BibitemShut
  {NoStop}%
\bibitem [{\citenamefont {Anisimov}\ \emph {et~al.}(2014)\citenamefont
  {Anisimov}, \citenamefont {Bauer}, \citenamefont {Chadalavada}, \citenamefont
  {Olson}, \citenamefont {Glenski}, \citenamefont {Kramer}, \citenamefont
  {Aprà},\ and\ \citenamefont {Kowalski}}]{Anisimov2014}%
  \BibitemOpen
  \bibfield  {author} {\bibinfo {author} {\bibfnamefont {V.~M.}\ \bibnamefont
  {Anisimov}}, \bibinfo {author} {\bibfnamefont {G.~H.}\ \bibnamefont {Bauer}},
  \bibinfo {author} {\bibfnamefont {K.}~\bibnamefont {Chadalavada}}, \bibinfo
  {author} {\bibfnamefont {R.~M.}\ \bibnamefont {Olson}}, \bibinfo {author}
  {\bibfnamefont {J.~W.}\ \bibnamefont {Glenski}}, \bibinfo {author}
  {\bibfnamefont {W.~T.~C.}\ \bibnamefont {Kramer}}, \bibinfo {author}
  {\bibfnamefont {E.}~\bibnamefont {Aprà}}, \ and\ \bibinfo {author}
  {\bibfnamefont {K.}~\bibnamefont {Kowalski}},\ }\href {\doibase
  10.1021/ct500404c} {\bibfield  {journal} {\bibinfo  {journal} {Journal of
  Chemical Theory and Computation}\ }\textbf {\bibinfo {volume} {10}},\
  \bibinfo {pages} {4307} (\bibinfo {year} {2014})}\BibitemShut {NoStop}%
\bibitem [{\citenamefont {Solomonik}\ \emph {et~al.}(2014)\citenamefont
  {Solomonik}, \citenamefont {Matthews}, \citenamefont {Hammond}, \citenamefont
  {Stanton},\ and\ \citenamefont {Demmel}}]{Solomonik2014}%
  \BibitemOpen
  \bibfield  {author} {\bibinfo {author} {\bibfnamefont {E.}~\bibnamefont
  {Solomonik}}, \bibinfo {author} {\bibfnamefont {D.}~\bibnamefont {Matthews}},
  \bibinfo {author} {\bibfnamefont {J.~R.}\ \bibnamefont {Hammond}}, \bibinfo
  {author} {\bibfnamefont {J.~F.}\ \bibnamefont {Stanton}}, \ and\ \bibinfo
  {author} {\bibfnamefont {J.}~\bibnamefont {Demmel}},\ }\href {\doibase
  10.1016/j.jpdc.2014.06.002} {\bibfield  {journal} {\bibinfo  {journal}
  {Journal of Parallel and Distributed Computing}\ }\textbf {\bibinfo {volume}
  {74}},\ \bibinfo {pages} {3176} (\bibinfo {year} {2014})}\BibitemShut
  {NoStop}%
\bibitem [{\citenamefont {Peng}\ \emph {et~al.}(2016)\citenamefont {Peng},
  \citenamefont {Calvin}, \citenamefont {Pavošević}, \citenamefont {Zhang},\
  and\ \citenamefont {Valeev}}]{Peng2016}%
  \BibitemOpen
  \bibfield  {author} {\bibinfo {author} {\bibfnamefont {C.}~\bibnamefont
  {Peng}}, \bibinfo {author} {\bibfnamefont {J.~A.}\ \bibnamefont {Calvin}},
  \bibinfo {author} {\bibfnamefont {F.}~\bibnamefont {Pavošević}}, \bibinfo
  {author} {\bibfnamefont {J.}~\bibnamefont {Zhang}}, \ and\ \bibinfo {author}
  {\bibfnamefont {E.~F.}\ \bibnamefont {Valeev}},\ }\href {\doibase
  10.1021/acs.jpca.6b10150} {\bibfield  {journal} {\bibinfo  {journal} {The
  Journal of Physical Chemistry A}\ }\textbf {\bibinfo {volume} {120}},\
  \bibinfo {pages} {10231} (\bibinfo {year} {2016})}\BibitemShut {NoStop}%
\bibitem [{\citenamefont {Lyakh}(2019)}]{Lyakh2019}%
  \BibitemOpen
  \bibfield  {author} {\bibinfo {author} {\bibfnamefont {D.~I.}\ \bibnamefont
  {Lyakh}},\ }\href {\doibase 10.1002/qua.25926} {\bibfield  {journal}
  {\bibinfo  {journal} {International Journal of Quantum Chemistry}\ }\textbf
  {\bibinfo {volume} {119}},\ \bibinfo {pages} {e25926} (\bibinfo {year}
  {2019})}\BibitemShut {NoStop}%
\bibitem [{\citenamefont {Gyevi-Nagy}\ \emph {et~al.}(2020)\citenamefont
  {Gyevi-Nagy}, \citenamefont {Kállay},\ and\ \citenamefont
  {Nagy}}]{Gyevi2020}%
  \BibitemOpen
  \bibfield  {author} {\bibinfo {author} {\bibfnamefont {L.}~\bibnamefont
  {Gyevi-Nagy}}, \bibinfo {author} {\bibfnamefont {M.}~\bibnamefont {Kállay}},
  \ and\ \bibinfo {author} {\bibfnamefont {P.~R.}\ \bibnamefont {Nagy}},\
  }\href {\doibase 10.1021/acs.jctc.9b00957} {\bibfield  {journal} {\bibinfo
  {journal} {Journal of Chemical Theory and Computation}\ }\textbf {\bibinfo
  {volume} {16}},\ \bibinfo {pages} {366} (\bibinfo {year} {2020})}\BibitemShut
  {NoStop}%
\bibitem [{\citenamefont {Peng}\ \emph {et~al.}(2020)\citenamefont {Peng},
  \citenamefont {Lewis}, \citenamefont {Wang}, \citenamefont {Clement},
  \citenamefont {Pierce}, \citenamefont {Rishi}, \citenamefont {Pavošević},
  \citenamefont {Slattery}, \citenamefont {Zhang}, \citenamefont {Teke},
  \citenamefont {Kumar}, \citenamefont {Masteran}, \citenamefont {Asadchev},
  \citenamefont {Calvin},\ and\ \citenamefont {Valeev}}]{Peng2020}%
  \BibitemOpen
  \bibfield  {author} {\bibinfo {author} {\bibfnamefont {C.}~\bibnamefont
  {Peng}}, \bibinfo {author} {\bibfnamefont {C.~A.}\ \bibnamefont {Lewis}},
  \bibinfo {author} {\bibfnamefont {X.}~\bibnamefont {Wang}}, \bibinfo {author}
  {\bibfnamefont {M.~C.}\ \bibnamefont {Clement}}, \bibinfo {author}
  {\bibfnamefont {K.}~\bibnamefont {Pierce}}, \bibinfo {author} {\bibfnamefont
  {V.}~\bibnamefont {Rishi}}, \bibinfo {author} {\bibfnamefont
  {F.}~\bibnamefont {Pavošević}}, \bibinfo {author} {\bibfnamefont
  {S.}~\bibnamefont {Slattery}}, \bibinfo {author} {\bibfnamefont
  {J.}~\bibnamefont {Zhang}}, \bibinfo {author} {\bibfnamefont
  {N.}~\bibnamefont {Teke}}, \bibinfo {author} {\bibfnamefont {A.}~\bibnamefont
  {Kumar}}, \bibinfo {author} {\bibfnamefont {C.}~\bibnamefont {Masteran}},
  \bibinfo {author} {\bibfnamefont {A.}~\bibnamefont {Asadchev}}, \bibinfo
  {author} {\bibfnamefont {J.~A.}\ \bibnamefont {Calvin}}, \ and\ \bibinfo
  {author} {\bibfnamefont {E.~F.}\ \bibnamefont {Valeev}},\ }\href {\doibase
  10.1063/5.0005889} {\bibfield  {journal} {\bibinfo  {journal} {The Journal of
  Chemical Physics}\ }\textbf {\bibinfo {volume} {153}},\ \bibinfo {pages}
  {044120} (\bibinfo {year} {2020})}\BibitemShut {NoStop}%
\bibitem [{\citenamefont {Datta}\ and\ \citenamefont
  {Gordon}(2021)}]{Datta2021}%
  \BibitemOpen
  \bibfield  {author} {\bibinfo {author} {\bibfnamefont {D.}~\bibnamefont
  {Datta}}\ and\ \bibinfo {author} {\bibfnamefont {M.~S.}\ \bibnamefont
  {Gordon}},\ }\href {\doibase 10.1021/acs.jctc.1c00389} {\bibfield  {journal}
  {\bibinfo  {journal} {Journal of Chemical Theory and Computation}\ }\textbf
  {\bibinfo {volume} {17}},\ \bibinfo {pages} {4799} (\bibinfo {year}
  {2021})}\BibitemShut {NoStop}%
\bibitem [{\citenamefont {Gyevi-Nagy}\ \emph {et~al.}(2021)\citenamefont
  {Gyevi-Nagy}, \citenamefont {Kállay},\ and\ \citenamefont
  {Nagy}}]{Gyevi2021}%
  \BibitemOpen
  \bibfield  {author} {\bibinfo {author} {\bibfnamefont {L.}~\bibnamefont
  {Gyevi-Nagy}}, \bibinfo {author} {\bibfnamefont {M.}~\bibnamefont {Kállay}},
  \ and\ \bibinfo {author} {\bibfnamefont {P.~R.}\ \bibnamefont {Nagy}},\
  }\href {\doibase 10.1021/acs.jctc.0c01077} {\bibfield  {journal} {\bibinfo
  {journal} {Journal of Chemical Theory and Computation}\ }\textbf {\bibinfo
  {volume} {17}},\ \bibinfo {pages} {860} (\bibinfo {year} {2021})}\BibitemShut
  {NoStop}%
\bibitem [{\citenamefont {Kowalski}\ \emph {et~al.}(2021)\citenamefont
  {Kowalski}, \citenamefont {Bair}, \citenamefont {Bauman}, \citenamefont
  {Boschen}, \citenamefont {Bylaska}, \citenamefont {Daily}, \citenamefont
  {de~Jong}, \citenamefont {Dunning}, \citenamefont {Govind}, \citenamefont
  {Harrison}, \citenamefont {Keçeli}, \citenamefont {Keipert}, \citenamefont
  {Krishnamoorthy}, \citenamefont {Kumar}, \citenamefont {Mutlu}, \citenamefont
  {Palmer}, \citenamefont {Panyala}, \citenamefont {Peng}, \citenamefont
  {Richard}, \citenamefont {Straatsma}, \citenamefont {Sushko}, \citenamefont
  {Valeev}, \citenamefont {Valiev}, \citenamefont {van Dam}, \citenamefont
  {Waldrop}, \citenamefont {Williams-Young}, \citenamefont {Yang},
  \citenamefont {Zalewski},\ and\ \citenamefont {Windus}}]{Kowalski2021}%
  \BibitemOpen
  \bibfield  {author} {\bibinfo {author} {\bibfnamefont {K.}~\bibnamefont
  {Kowalski}}, \bibinfo {author} {\bibfnamefont {R.}~\bibnamefont {Bair}},
  \bibinfo {author} {\bibfnamefont {N.~P.}\ \bibnamefont {Bauman}}, \bibinfo
  {author} {\bibfnamefont {J.~S.}\ \bibnamefont {Boschen}}, \bibinfo {author}
  {\bibfnamefont {E.~J.}\ \bibnamefont {Bylaska}}, \bibinfo {author}
  {\bibfnamefont {J.}~\bibnamefont {Daily}}, \bibinfo {author} {\bibfnamefont
  {W.~A.}\ \bibnamefont {de~Jong}}, \bibinfo {author} {\bibfnamefont
  {T.}~\bibnamefont {Dunning}}, \bibinfo {author} {\bibfnamefont
  {N.}~\bibnamefont {Govind}}, \bibinfo {author} {\bibfnamefont {R.~J.}\
  \bibnamefont {Harrison}}, \bibinfo {author} {\bibfnamefont {M.}~\bibnamefont
  {Keçeli}}, \bibinfo {author} {\bibfnamefont {K.}~\bibnamefont {Keipert}},
  \bibinfo {author} {\bibfnamefont {S.}~\bibnamefont {Krishnamoorthy}},
  \bibinfo {author} {\bibfnamefont {S.}~\bibnamefont {Kumar}}, \bibinfo
  {author} {\bibfnamefont {E.}~\bibnamefont {Mutlu}}, \bibinfo {author}
  {\bibfnamefont {B.}~\bibnamefont {Palmer}}, \bibinfo {author} {\bibfnamefont
  {A.}~\bibnamefont {Panyala}}, \bibinfo {author} {\bibfnamefont
  {B.}~\bibnamefont {Peng}}, \bibinfo {author} {\bibfnamefont {R.~M.}\
  \bibnamefont {Richard}}, \bibinfo {author} {\bibfnamefont {T.~P.}\
  \bibnamefont {Straatsma}}, \bibinfo {author} {\bibfnamefont {P.}~\bibnamefont
  {Sushko}}, \bibinfo {author} {\bibfnamefont {E.~F.}\ \bibnamefont {Valeev}},
  \bibinfo {author} {\bibfnamefont {M.}~\bibnamefont {Valiev}}, \bibinfo
  {author} {\bibfnamefont {H.~J.~J.}\ \bibnamefont {van Dam}}, \bibinfo
  {author} {\bibfnamefont {J.~M.}\ \bibnamefont {Waldrop}}, \bibinfo {author}
  {\bibfnamefont {D.~B.}\ \bibnamefont {Williams-Young}}, \bibinfo {author}
  {\bibfnamefont {C.}~\bibnamefont {Yang}}, \bibinfo {author} {\bibfnamefont
  {M.}~\bibnamefont {Zalewski}}, \ and\ \bibinfo {author} {\bibfnamefont
  {T.~L.}\ \bibnamefont {Windus}},\ }\href {\doibase
  10.1021/acs.chemrev.0c00998} {\bibfield  {journal} {\bibinfo  {journal}
  {Chemical Reviews}\ }\textbf {\bibinfo {volume} {121}},\ \bibinfo {pages}
  {4962} (\bibinfo {year} {2021})}\BibitemShut {NoStop}%
\bibitem [{\citenamefont {Calvin}\ \emph {et~al.}(2021)\citenamefont {Calvin},
  \citenamefont {Peng}, \citenamefont {Rishi}, \citenamefont {Kumar},\ and\
  \citenamefont {Valeev}}]{Calvin2021}%
  \BibitemOpen
  \bibfield  {author} {\bibinfo {author} {\bibfnamefont {J.~A.}\ \bibnamefont
  {Calvin}}, \bibinfo {author} {\bibfnamefont {C.}~\bibnamefont {Peng}},
  \bibinfo {author} {\bibfnamefont {V.}~\bibnamefont {Rishi}}, \bibinfo
  {author} {\bibfnamefont {A.}~\bibnamefont {Kumar}}, \ and\ \bibinfo {author}
  {\bibfnamefont {E.~F.}\ \bibnamefont {Valeev}},\ }\href {\doibase
  10.1021/acs.chemrev.0c00006} {\bibfield  {journal} {\bibinfo  {journal}
  {Chemical Reviews}\ }\textbf {\bibinfo {volume} {121}},\ \bibinfo {pages}
  {1203} (\bibinfo {year} {2021})}\BibitemShut {NoStop}%
\bibitem [{\citenamefont {Seritan}\ \emph {et~al.}(2020)\citenamefont
  {Seritan}, \citenamefont {Bannwarth}, \citenamefont {Fales}, \citenamefont
  {Hohenstein}, \citenamefont {Kokkila-Schumacher}, \citenamefont {Luehr},
  \citenamefont {Snyder}, \citenamefont {Song}, \citenamefont {Titov},
  \citenamefont {Ufimtsev},\ and\ \citenamefont {Martínez}}]{Seritan2020}%
  \BibitemOpen
  \bibfield  {author} {\bibinfo {author} {\bibfnamefont {S.}~\bibnamefont
  {Seritan}}, \bibinfo {author} {\bibfnamefont {C.}~\bibnamefont {Bannwarth}},
  \bibinfo {author} {\bibfnamefont {B.~S.}\ \bibnamefont {Fales}}, \bibinfo
  {author} {\bibfnamefont {E.~G.}\ \bibnamefont {Hohenstein}}, \bibinfo
  {author} {\bibfnamefont {S.~I.~L.}\ \bibnamefont {Kokkila-Schumacher}},
  \bibinfo {author} {\bibfnamefont {N.}~\bibnamefont {Luehr}}, \bibinfo
  {author} {\bibfnamefont {J.~W.}\ \bibnamefont {Snyder}}, \bibinfo {author}
  {\bibfnamefont {C.}~\bibnamefont {Song}}, \bibinfo {author} {\bibfnamefont
  {A.~V.}\ \bibnamefont {Titov}}, \bibinfo {author} {\bibfnamefont {I.~S.}\
  \bibnamefont {Ufimtsev}}, \ and\ \bibinfo {author} {\bibfnamefont {T.~J.}\
  \bibnamefont {Martínez}},\ }\href {\doibase 10.1063/5.0007615} {\bibfield
  {journal} {\bibinfo  {journal} {The Journal of Chemical Physics}\ }\textbf
  {\bibinfo {volume} {152}},\ \bibinfo {pages} {224110} (\bibinfo {year}
  {2020})}\BibitemShut {NoStop}%
\bibitem [{\citenamefont {Wang}\ \emph {et~al.}(2020)\citenamefont {Wang},
  \citenamefont {Guo},\ and\ \citenamefont {Wang}}]{Wang2020}%
  \BibitemOpen
  \bibfield  {author} {\bibinfo {author} {\bibfnamefont {Z.}~\bibnamefont
  {Wang}}, \bibinfo {author} {\bibfnamefont {M.}~\bibnamefont {Guo}}, \ and\
  \bibinfo {author} {\bibfnamefont {F.}~\bibnamefont {Wang}},\ }\href {\doibase
  10.1039/D0CP03800H} {\bibfield  {journal} {\bibinfo  {journal} {Physical
  Chemistry Chemical Physics}\ }\textbf {\bibinfo {volume} {22}},\ \bibinfo
  {pages} {25103} (\bibinfo {year} {2020})}\BibitemShut {NoStop}%
\bibitem [{\citenamefont {Peng}\ \emph {et~al.}(2019)\citenamefont {Peng},
  \citenamefont {Calvin},\ and\ \citenamefont {Valeev}}]{Peng2019}%
  \BibitemOpen
  \bibfield  {author} {\bibinfo {author} {\bibfnamefont {C.}~\bibnamefont
  {Peng}}, \bibinfo {author} {\bibfnamefont {J.~A.}\ \bibnamefont {Calvin}}, \
  and\ \bibinfo {author} {\bibfnamefont {E.~F.}\ \bibnamefont {Valeev}},\
  }\href {\doibase 10.1002/qua.25894} {\bibfield  {journal} {\bibinfo
  {journal} {International Journal of Quantum Chemistry}\ }\textbf {\bibinfo
  {volume} {119}} (\bibinfo {year} {2019}),\ 10.1002/qua.25894}\BibitemShut
  {NoStop}%
\bibitem [{\citenamefont {Kaliman}\ and\ \citenamefont
  {Krylov}(2017)}]{Kaliman2017}%
  \BibitemOpen
  \bibfield  {author} {\bibinfo {author} {\bibfnamefont {I.~A.}\ \bibnamefont
  {Kaliman}}\ and\ \bibinfo {author} {\bibfnamefont {A.~I.}\ \bibnamefont
  {Krylov}},\ }\href {\doibase 10.1002/jcc.24713} {\bibfield  {journal}
  {\bibinfo  {journal} {Journal of Computational Chemistry}\ }\textbf {\bibinfo
  {volume} {38}},\ \bibinfo {pages} {842} (\bibinfo {year} {2017})}\BibitemShut
  {NoStop}%
\bibitem [{\citenamefont {DePrince}\ \emph {et~al.}(2014)\citenamefont
  {DePrince}, \citenamefont {Kennedy}, \citenamefont {Sumpter},\ and\
  \citenamefont {Sherrill}}]{DePrince2014}%
  \BibitemOpen
  \bibfield  {author} {\bibinfo {author} {\bibfnamefont {A.~E.}\ \bibnamefont
  {DePrince}}, \bibinfo {author} {\bibfnamefont {M.~R.}\ \bibnamefont
  {Kennedy}}, \bibinfo {author} {\bibfnamefont {B.~G.}\ \bibnamefont
  {Sumpter}}, \ and\ \bibinfo {author} {\bibfnamefont {C.~D.}\ \bibnamefont
  {Sherrill}},\ }\href {\doibase 10.1080/00268976.2013.874599} {\bibfield
  {journal} {\bibinfo  {journal} {Molecular Physics}\ }\textbf {\bibinfo
  {volume} {112}},\ \bibinfo {pages} {844} (\bibinfo {year}
  {2014})}\BibitemShut {NoStop}%
\bibitem [{\citenamefont {Ma}\ \emph {et~al.}(2011)\citenamefont {Ma},
  \citenamefont {Krishnamoorthy}, \citenamefont {Villa},\ and\ \citenamefont
  {Kowalski}}]{Ma2011}%
  \BibitemOpen
  \bibfield  {author} {\bibinfo {author} {\bibfnamefont {W.}~\bibnamefont
  {Ma}}, \bibinfo {author} {\bibfnamefont {S.}~\bibnamefont {Krishnamoorthy}},
  \bibinfo {author} {\bibfnamefont {O.}~\bibnamefont {Villa}}, \ and\ \bibinfo
  {author} {\bibfnamefont {K.}~\bibnamefont {Kowalski}},\ }\href {\doibase
  10.1021/ct1007247} {\bibfield  {journal} {\bibinfo  {journal} {Journal of
  Chemical Theory and Computation}\ }\textbf {\bibinfo {volume} {7}},\ \bibinfo
  {pages} {1316} (\bibinfo {year} {2011})}\BibitemShut {NoStop}%
\bibitem [{\citenamefont {DePrince}\ and\ \citenamefont
  {Hammond}(2011)}]{DePrince2011}%
  \BibitemOpen
  \bibfield  {author} {\bibinfo {author} {\bibfnamefont {A.~E.}\ \bibnamefont
  {DePrince}}\ and\ \bibinfo {author} {\bibfnamefont {J.~R.}\ \bibnamefont
  {Hammond}},\ }\href {\doibase 10.1021/ct100584w} {\bibfield  {journal}
  {\bibinfo  {journal} {Journal of Chemical Theory and Computation}\ }\textbf
  {\bibinfo {volume} {7}},\ \bibinfo {pages} {1287} (\bibinfo {year}
  {2011})}\BibitemShut {NoStop}%
\bibitem [{\citenamefont {Møller}\ and\ \citenamefont
  {Plesset}(1934)}]{Moller1934}%
  \BibitemOpen
  \bibfield  {author} {\bibinfo {author} {\bibfnamefont {C.}~\bibnamefont
  {Møller}}\ and\ \bibinfo {author} {\bibfnamefont {M.~S.}\ \bibnamefont
  {Plesset}},\ }\href {\doibase 10.1103/PhysRev.46.618} {\bibfield  {journal}
  {\bibinfo  {journal} {Physical Review}\ }\textbf {\bibinfo {volume} {46}},\
  \bibinfo {pages} {618} (\bibinfo {year} {1934})}\BibitemShut {NoStop}%
\bibitem [{\citenamefont {Cremer}(2011)}]{Cremer2011}%
  \BibitemOpen
  \bibfield  {author} {\bibinfo {author} {\bibfnamefont {D.}~\bibnamefont
  {Cremer}},\ }\href {\doibase 10.1002/wcms.58} {\bibfield  {journal} {\bibinfo
   {journal} {Wiley Interdisciplinary Reviews: Computational Molecular
  Science}\ }\textbf {\bibinfo {volume} {1}},\ \bibinfo {pages} {509} (\bibinfo
  {year} {2011})}\BibitemShut {NoStop}%
\bibitem [{\citenamefont {Hohenberg}\ and\ \citenamefont
  {Kohn}(1964)}]{Hohenberg1964}%
  \BibitemOpen
  \bibfield  {author} {\bibinfo {author} {\bibfnamefont {P.}~\bibnamefont
  {Hohenberg}}\ and\ \bibinfo {author} {\bibfnamefont {W.}~\bibnamefont
  {Kohn}},\ }\href {\doibase 10.1103/PhysRev.136.B864} {\bibfield  {journal}
  {\bibinfo  {journal} {Physical Review}\ }\textbf {\bibinfo {volume} {136}},\
  \bibinfo {pages} {B864} (\bibinfo {year} {1964})}\BibitemShut {NoStop}%
\bibitem [{\citenamefont {Kohn}\ and\ \citenamefont {Sham}(1965)}]{Kohn1965}%
  \BibitemOpen
  \bibfield  {author} {\bibinfo {author} {\bibfnamefont {W.}~\bibnamefont
  {Kohn}}\ and\ \bibinfo {author} {\bibfnamefont {L.~J.}\ \bibnamefont
  {Sham}},\ }\href {\doibase 10.1103/PhysRev.140.A1133} {\bibfield  {journal}
  {\bibinfo  {journal} {Physical Review}\ }\textbf {\bibinfo {volume} {140}},\
  \bibinfo {pages} {A1133} (\bibinfo {year} {1965})}\BibitemShut {NoStop}%
\bibitem [{\citenamefont {Maurer}\ \emph {et~al.}(2014)\citenamefont {Maurer},
  \citenamefont {Clin},\ and\ \citenamefont {Ochsenfeld}}]{Maurer2014}%
  \BibitemOpen
  \bibfield  {author} {\bibinfo {author} {\bibfnamefont {S.~A.}\ \bibnamefont
  {Maurer}}, \bibinfo {author} {\bibfnamefont {L.}~\bibnamefont {Clin}}, \ and\
  \bibinfo {author} {\bibfnamefont {C.}~\bibnamefont {Ochsenfeld}},\ }\href
  {\doibase 10.1063/1.4881144} {\bibfield  {journal} {\bibinfo  {journal} {The
  Journal of Chemical Physics}\ }\textbf {\bibinfo {volume} {140}},\ \bibinfo
  {pages} {224112} (\bibinfo {year} {2014})}\BibitemShut {NoStop}%
\bibitem [{\citenamefont {Dawson}\ \emph {et~al.}(2022)\citenamefont {Dawson},
  \citenamefont {Degomme}, \citenamefont {Stella}, \citenamefont {Nakajima},
  \citenamefont {Ratcliff},\ and\ \citenamefont {Genovese}}]{Dawson2022}%
  \BibitemOpen
  \bibfield  {author} {\bibinfo {author} {\bibfnamefont {W.}~\bibnamefont
  {Dawson}}, \bibinfo {author} {\bibfnamefont {A.}~\bibnamefont {Degomme}},
  \bibinfo {author} {\bibfnamefont {M.}~\bibnamefont {Stella}}, \bibinfo
  {author} {\bibfnamefont {T.}~\bibnamefont {Nakajima}}, \bibinfo {author}
  {\bibfnamefont {L.~E.}\ \bibnamefont {Ratcliff}}, \ and\ \bibinfo {author}
  {\bibfnamefont {L.}~\bibnamefont {Genovese}},\ }\href {\doibase
  10.1002/wcms.1574} {\bibfield  {journal} {\bibinfo  {journal} {WIREs
  Computational Molecular Science}\ }\textbf {\bibinfo {volume} {12}} (\bibinfo
  {year} {2022}),\ 10.1002/wcms.1574}\BibitemShut {NoStop}%
\bibitem [{\citenamefont {Li}\ \emph {et~al.}(2002)\citenamefont {Li},
  \citenamefont {Ma},\ and\ \citenamefont {Jiang}}]{Li2002}%
  \BibitemOpen
  \bibfield  {author} {\bibinfo {author} {\bibfnamefont {S.}~\bibnamefont
  {Li}}, \bibinfo {author} {\bibfnamefont {J.}~\bibnamefont {Ma}}, \ and\
  \bibinfo {author} {\bibfnamefont {Y.}~\bibnamefont {Jiang}},\ }\href
  {\doibase 10.1002/jcc.10003} {\bibfield  {journal} {\bibinfo  {journal}
  {Journal of Computational Chemistry}\ }\textbf {\bibinfo {volume} {23}},\
  \bibinfo {pages} {237} (\bibinfo {year} {2002})}\BibitemShut {NoStop}%
\bibitem [{\citenamefont {Li}\ \emph {et~al.}(2006)\citenamefont {Li},
  \citenamefont {Shen}, \citenamefont {Li},\ and\ \citenamefont
  {Jiang}}]{Li2006}%
  \BibitemOpen
  \bibfield  {author} {\bibinfo {author} {\bibfnamefont {S.}~\bibnamefont
  {Li}}, \bibinfo {author} {\bibfnamefont {J.}~\bibnamefont {Shen}}, \bibinfo
  {author} {\bibfnamefont {W.}~\bibnamefont {Li}}, \ and\ \bibinfo {author}
  {\bibfnamefont {Y.}~\bibnamefont {Jiang}},\ }\href {\doibase
  10.1063/1.2244566} {\bibfield  {journal} {\bibinfo  {journal} {The Journal of
  Chemical Physics}\ }\textbf {\bibinfo {volume} {125}},\ \bibinfo {pages}
  {074109} (\bibinfo {year} {2006})}\BibitemShut {NoStop}%
\bibitem [{\citenamefont {Li}\ \emph {et~al.}(2009)\citenamefont {Li},
  \citenamefont {Piecuch}, \citenamefont {Gour},\ and\ \citenamefont
  {Li}}]{Li2009}%
  \BibitemOpen
  \bibfield  {author} {\bibinfo {author} {\bibfnamefont {W.}~\bibnamefont
  {Li}}, \bibinfo {author} {\bibfnamefont {P.}~\bibnamefont {Piecuch}},
  \bibinfo {author} {\bibfnamefont {J.~R.}\ \bibnamefont {Gour}}, \ and\
  \bibinfo {author} {\bibfnamefont {S.}~\bibnamefont {Li}},\ }\href {\doibase
  10.1063/1.3218842} {\bibfield  {journal} {\bibinfo  {journal} {The Journal of
  Chemical Physics}\ }\textbf {\bibinfo {volume} {131}},\ \bibinfo {pages}
  {114109} (\bibinfo {year} {2009})}\BibitemShut {NoStop}%
\bibitem [{\citenamefont {Neese}\ \emph {et~al.}(2009)\citenamefont {Neese},
  \citenamefont {Wennmohs},\ and\ \citenamefont {Hansen}}]{Neese2009}%
  \BibitemOpen
  \bibfield  {author} {\bibinfo {author} {\bibfnamefont {F.}~\bibnamefont
  {Neese}}, \bibinfo {author} {\bibfnamefont {F.}~\bibnamefont {Wennmohs}}, \
  and\ \bibinfo {author} {\bibfnamefont {A.}~\bibnamefont {Hansen}},\ }\href
  {\doibase 10.1063/1.3086717} {\bibfield  {journal} {\bibinfo  {journal} {The
  Journal of Chemical Physics}\ }\textbf {\bibinfo {volume} {130}},\ \bibinfo
  {pages} {114108} (\bibinfo {year} {2009})}\BibitemShut {NoStop}%
\bibitem [{\citenamefont {Li}\ and\ \citenamefont
  {Piecuch}(2010{\natexlab{a}})}]{Li2010a}%
  \BibitemOpen
  \bibfield  {author} {\bibinfo {author} {\bibfnamefont {W.}~\bibnamefont
  {Li}}\ and\ \bibinfo {author} {\bibfnamefont {P.}~\bibnamefont {Piecuch}},\
  }\href {\doibase 10.1021/jp1038738} {\bibfield  {journal} {\bibinfo
  {journal} {The Journal of Physical Chemistry A}\ }\textbf {\bibinfo {volume}
  {114}},\ \bibinfo {pages} {6721} (\bibinfo {year}
  {2010}{\natexlab{a}})}\BibitemShut {NoStop}%
\bibitem [{\citenamefont {Li}\ and\ \citenamefont
  {Piecuch}(2010{\natexlab{b}})}]{Li2010b}%
  \BibitemOpen
  \bibfield  {author} {\bibinfo {author} {\bibfnamefont {W.}~\bibnamefont
  {Li}}\ and\ \bibinfo {author} {\bibfnamefont {P.}~\bibnamefont {Piecuch}},\
  }\href {\doibase 10.1021/jp100782u} {\bibfield  {journal} {\bibinfo
  {journal} {The Journal of Physical Chemistry A}\ }\textbf {\bibinfo {volume}
  {114}},\ \bibinfo {pages} {8644} (\bibinfo {year}
  {2010}{\natexlab{b}})}\BibitemShut {NoStop}%
\bibitem [{\citenamefont {Rolik}\ and\ \citenamefont
  {Kállay}(2011)}]{Rolik2011}%
  \BibitemOpen
  \bibfield  {author} {\bibinfo {author} {\bibfnamefont {Z.}~\bibnamefont
  {Rolik}}\ and\ \bibinfo {author} {\bibfnamefont {M.}~\bibnamefont
  {Kállay}},\ }\href {\doibase 10.1063/1.3632085} {\bibfield  {journal}
  {\bibinfo  {journal} {The Journal of Chemical Physics}\ }\textbf {\bibinfo
  {volume} {135}},\ \bibinfo {pages} {104111} (\bibinfo {year}
  {2011})}\BibitemShut {NoStop}%
\bibitem [{\citenamefont {Rolik}\ \emph {et~al.}(2013)\citenamefont {Rolik},
  \citenamefont {Szegedy}, \citenamefont {Ladjánszki}, \citenamefont
  {Ladóczki},\ and\ \citenamefont {Kállay}}]{Rolik2013}%
  \BibitemOpen
  \bibfield  {author} {\bibinfo {author} {\bibfnamefont {Z.}~\bibnamefont
  {Rolik}}, \bibinfo {author} {\bibfnamefont {L.}~\bibnamefont {Szegedy}},
  \bibinfo {author} {\bibfnamefont {I.}~\bibnamefont {Ladjánszki}}, \bibinfo
  {author} {\bibfnamefont {B.}~\bibnamefont {Ladóczki}}, \ and\ \bibinfo
  {author} {\bibfnamefont {M.}~\bibnamefont {Kállay}},\ }\href {\doibase
  10.1063/1.4819401} {\bibfield  {journal} {\bibinfo  {journal} {The Journal of
  Chemical Physics}\ }\textbf {\bibinfo {volume} {139}},\ \bibinfo {pages}
  {094105} (\bibinfo {year} {2013})}\BibitemShut {NoStop}%
\bibitem [{\citenamefont {Riplinger}\ and\ \citenamefont
  {Neese}(2013)}]{Riplinger2013a}%
  \BibitemOpen
  \bibfield  {author} {\bibinfo {author} {\bibfnamefont {C.}~\bibnamefont
  {Riplinger}}\ and\ \bibinfo {author} {\bibfnamefont {F.}~\bibnamefont
  {Neese}},\ }\href {\doibase 10.1063/1.4773581} {\bibfield  {journal}
  {\bibinfo  {journal} {The Journal of Chemical Physics}\ }\textbf {\bibinfo
  {volume} {138}},\ \bibinfo {pages} {034106} (\bibinfo {year}
  {2013})}\BibitemShut {NoStop}%
\bibitem [{\citenamefont {Riplinger}\ \emph {et~al.}(2013)\citenamefont
  {Riplinger}, \citenamefont {Sandhoefer}, \citenamefont {Hansen},\ and\
  \citenamefont {Neese}}]{Riplinger2013b}%
  \BibitemOpen
  \bibfield  {author} {\bibinfo {author} {\bibfnamefont {C.}~\bibnamefont
  {Riplinger}}, \bibinfo {author} {\bibfnamefont {B.}~\bibnamefont
  {Sandhoefer}}, \bibinfo {author} {\bibfnamefont {A.}~\bibnamefont {Hansen}},
  \ and\ \bibinfo {author} {\bibfnamefont {F.}~\bibnamefont {Neese}},\ }\href
  {\doibase 10.1063/1.4821834} {\bibfield  {journal} {\bibinfo  {journal} {The
  Journal of Chemical Physics}\ }\textbf {\bibinfo {volume} {139}},\ \bibinfo
  {pages} {134101} (\bibinfo {year} {2013})}\BibitemShut {NoStop}%
\bibitem [{\citenamefont {Liakos}\ \emph {et~al.}(2015)\citenamefont {Liakos},
  \citenamefont {Sparta}, \citenamefont {Kesharwani}, \citenamefont {Martin},\
  and\ \citenamefont {Neese}}]{Liakos2015}%
  \BibitemOpen
  \bibfield  {author} {\bibinfo {author} {\bibfnamefont {D.~G.}\ \bibnamefont
  {Liakos}}, \bibinfo {author} {\bibfnamefont {M.}~\bibnamefont {Sparta}},
  \bibinfo {author} {\bibfnamefont {M.~K.}\ \bibnamefont {Kesharwani}},
  \bibinfo {author} {\bibfnamefont {J.~M.~L.}\ \bibnamefont {Martin}}, \ and\
  \bibinfo {author} {\bibfnamefont {F.}~\bibnamefont {Neese}},\ }\href
  {\doibase 10.1021/ct501129s} {\bibfield  {journal} {\bibinfo  {journal}
  {Journal of Chemical Theory and Computation}\ }\textbf {\bibinfo {volume}
  {11}},\ \bibinfo {pages} {1525} (\bibinfo {year} {2015})}\BibitemShut
  {NoStop}%
\bibitem [{\citenamefont {Schwilk}\ \emph {et~al.}(2017)\citenamefont
  {Schwilk}, \citenamefont {Ma}, \citenamefont {Köppl},\ and\ \citenamefont
  {Werner}}]{Schwilk2017}%
  \BibitemOpen
  \bibfield  {author} {\bibinfo {author} {\bibfnamefont {M.}~\bibnamefont
  {Schwilk}}, \bibinfo {author} {\bibfnamefont {Q.}~\bibnamefont {Ma}},
  \bibinfo {author} {\bibfnamefont {C.}~\bibnamefont {Köppl}}, \ and\ \bibinfo
  {author} {\bibfnamefont {H.-J.}\ \bibnamefont {Werner}},\ }\href {\doibase
  10.1021/acs.jctc.7b00554} {\bibfield  {journal} {\bibinfo  {journal} {Journal
  of Chemical Theory and Computation}\ }\textbf {\bibinfo {volume} {13}},\
  \bibinfo {pages} {3650} (\bibinfo {year} {2017})}\BibitemShut {NoStop}%
\bibitem [{\citenamefont {Pinski}\ \emph {et~al.}(2015)\citenamefont {Pinski},
  \citenamefont {Riplinger}, \citenamefont {Valeev},\ and\ \citenamefont
  {Neese}}]{Pinski2015}%
  \BibitemOpen
  \bibfield  {author} {\bibinfo {author} {\bibfnamefont {P.}~\bibnamefont
  {Pinski}}, \bibinfo {author} {\bibfnamefont {C.}~\bibnamefont {Riplinger}},
  \bibinfo {author} {\bibfnamefont {E.~F.}\ \bibnamefont {Valeev}}, \ and\
  \bibinfo {author} {\bibfnamefont {F.}~\bibnamefont {Neese}},\ }\href
  {\doibase 10.1063/1.4926879} {\bibfield  {journal} {\bibinfo  {journal} {The
  Journal of Chemical Physics}\ }\textbf {\bibinfo {volume} {143}},\ \bibinfo
  {pages} {034108} (\bibinfo {year} {2015})}\BibitemShut {NoStop}%
\bibitem [{\citenamefont {Riplinger}\ \emph {et~al.}(2016)\citenamefont
  {Riplinger}, \citenamefont {Pinski}, \citenamefont {Becker}, \citenamefont
  {Valeev},\ and\ \citenamefont {Neese}}]{Riplinger2016}%
  \BibitemOpen
  \bibfield  {author} {\bibinfo {author} {\bibfnamefont {C.}~\bibnamefont
  {Riplinger}}, \bibinfo {author} {\bibfnamefont {P.}~\bibnamefont {Pinski}},
  \bibinfo {author} {\bibfnamefont {U.}~\bibnamefont {Becker}}, \bibinfo
  {author} {\bibfnamefont {E.~F.}\ \bibnamefont {Valeev}}, \ and\ \bibinfo
  {author} {\bibfnamefont {F.}~\bibnamefont {Neese}},\ }\href {\doibase
  10.1063/1.4939030} {\bibfield  {journal} {\bibinfo  {journal} {The Journal of
  Chemical Physics}\ }\textbf {\bibinfo {volume} {144}},\ \bibinfo {pages}
  {024109} (\bibinfo {year} {2016})}\BibitemShut {NoStop}%
\bibitem [{\citenamefont {Parrish}\ \emph {et~al.}(2019)\citenamefont
  {Parrish}, \citenamefont {Zhao}, \citenamefont {Hohenstein},\ and\
  \citenamefont {Martínez}}]{Parrish2019}%
  \BibitemOpen
  \bibfield  {author} {\bibinfo {author} {\bibfnamefont {R.~M.}\ \bibnamefont
  {Parrish}}, \bibinfo {author} {\bibfnamefont {Y.}~\bibnamefont {Zhao}},
  \bibinfo {author} {\bibfnamefont {E.~G.}\ \bibnamefont {Hohenstein}}, \ and\
  \bibinfo {author} {\bibfnamefont {T.~J.}\ \bibnamefont {Martínez}},\ }\href
  {\doibase 10.1063/1.5092505} {\bibfield  {journal} {\bibinfo  {journal} {The
  Journal of Chemical Physics}\ }\textbf {\bibinfo {volume} {150}},\ \bibinfo
  {pages} {164118} (\bibinfo {year} {2019})},\ \Eprint
  {http://arxiv.org/abs/https://doi.org/10.1063/1.5092505}
  {https://doi.org/10.1063/1.5092505} \BibitemShut {NoStop}%
\bibitem [{\citenamefont {Kolda}\ and\ \citenamefont
  {Bader}(2009)}]{Kolda2009}%
  \BibitemOpen
  \bibfield  {author} {\bibinfo {author} {\bibfnamefont {T.~G.}\ \bibnamefont
  {Kolda}}\ and\ \bibinfo {author} {\bibfnamefont {B.~W.}\ \bibnamefont
  {Bader}},\ }\href {\doibase 10.1137/07070111X} {\bibfield  {journal}
  {\bibinfo  {journal} {SIAM Review}\ }\textbf {\bibinfo {volume} {51}},\
  \bibinfo {pages} {455} (\bibinfo {year} {2009})}\BibitemShut {NoStop}%
\bibitem [{\citenamefont {Hohenstein}\ \emph {et~al.}(2022)\citenamefont
  {Hohenstein}, \citenamefont {Fales}, \citenamefont {Parrish},\ and\
  \citenamefont {Martínez}}]{Hohenstein2022}%
  \BibitemOpen
  \bibfield  {author} {\bibinfo {author} {\bibfnamefont {E.~G.}\ \bibnamefont
  {Hohenstein}}, \bibinfo {author} {\bibfnamefont {B.~S.}\ \bibnamefont
  {Fales}}, \bibinfo {author} {\bibfnamefont {R.~M.}\ \bibnamefont {Parrish}},
  \ and\ \bibinfo {author} {\bibfnamefont {T.~J.}\ \bibnamefont {Martínez}},\
  }\href {\doibase 10.1063/5.0077770} {\bibfield  {journal} {\bibinfo
  {journal} {The Journal of Chemical Physics}\ }\textbf {\bibinfo {volume}
  {156}},\ \bibinfo {pages} {054102} (\bibinfo {year} {2022})},\ \Eprint
  {http://arxiv.org/abs/https://doi.org/10.1063/5.0077770}
  {https://doi.org/10.1063/5.0077770} \BibitemShut {NoStop}%
\bibitem [{\citenamefont {Lesiuk}(2020)}]{Lesiuk2019}%
  \BibitemOpen
  \bibfield  {author} {\bibinfo {author} {\bibfnamefont {M.}~\bibnamefont
  {Lesiuk}},\ }\href {\doibase 10.1021/acs.jctc.9b00985} {\bibfield  {journal}
  {\bibinfo  {journal} {Journal of Chemical Theory and Computation}\ }\textbf
  {\bibinfo {volume} {16}},\ \bibinfo {pages} {453} (\bibinfo {year} {2020})},\
  \bibinfo {note} {pMID: 31715103},\ \Eprint
  {http://arxiv.org/abs/https://doi.org/10.1021/acs.jctc.9b00985}
  {https://doi.org/10.1021/acs.jctc.9b00985} \BibitemShut {NoStop}%
\bibitem [{\citenamefont {Lesiuk}(2022)}]{Lesiuk2022}%
  \BibitemOpen
  \bibfield  {author} {\bibinfo {author} {\bibfnamefont {M.}~\bibnamefont
  {Lesiuk}},\ }\href {\doibase 10.1063/5.0071916} {\bibfield  {journal}
  {\bibinfo  {journal} {The Journal of Chemical Physics}\ }\textbf {\bibinfo
  {volume} {156}},\ \bibinfo {pages} {064103} (\bibinfo {year} {2022})},\
  \Eprint {http://arxiv.org/abs/https://doi.org/10.1063/5.0071916}
  {https://doi.org/10.1063/5.0071916} \BibitemShut {NoStop}%
\bibitem [{\citenamefont {Riley}\ \emph {et~al.}(2010)\citenamefont {Riley},
  \citenamefont {Pitoňák}, \citenamefont {Jurečka},\ and\ \citenamefont
  {Hobza}}]{Riley2010}%
  \BibitemOpen
  \bibfield  {author} {\bibinfo {author} {\bibfnamefont {K.~E.}\ \bibnamefont
  {Riley}}, \bibinfo {author} {\bibfnamefont {M.}~\bibnamefont {Pitoňák}},
  \bibinfo {author} {\bibfnamefont {P.}~\bibnamefont {Jurečka}}, \ and\
  \bibinfo {author} {\bibfnamefont {P.}~\bibnamefont {Hobza}},\ }\href
  {\doibase 10.1021/cr1000173} {\bibfield  {journal} {\bibinfo  {journal}
  {Chemical Reviews}\ }\textbf {\bibinfo {volume} {110}},\ \bibinfo {pages}
  {5023} (\bibinfo {year} {2010})}\BibitemShut {NoStop}%
\bibitem [{\citenamefont {Karton}\ \emph {et~al.}(2006)\citenamefont {Karton},
  \citenamefont {Rabinovich}, \citenamefont {Martin},\ and\ \citenamefont
  {Ruscic}}]{Karton2006}%
  \BibitemOpen
  \bibfield  {author} {\bibinfo {author} {\bibfnamefont {A.}~\bibnamefont
  {Karton}}, \bibinfo {author} {\bibfnamefont {E.}~\bibnamefont {Rabinovich}},
  \bibinfo {author} {\bibfnamefont {J.~M.~L.}\ \bibnamefont {Martin}}, \ and\
  \bibinfo {author} {\bibfnamefont {B.}~\bibnamefont {Ruscic}},\ }\href
  {\doibase 10.1063/1.2348881} {\bibfield  {journal} {\bibinfo  {journal} {The
  Journal of Chemical Physics}\ }\textbf {\bibinfo {volume} {125}},\ \bibinfo
  {pages} {144108} (\bibinfo {year} {2006})}\BibitemShut {NoStop}%
\bibitem [{\citenamefont {Tajti}\ \emph {et~al.}(2004)\citenamefont {Tajti},
  \citenamefont {Szalay}, \citenamefont {Császár}, \citenamefont {Kállay},
  \citenamefont {Gauss}, \citenamefont {Valeev}, \citenamefont {Flowers},
  \citenamefont {Vázquez},\ and\ \citenamefont {Stanton}}]{Tajti2004}%
  \BibitemOpen
  \bibfield  {author} {\bibinfo {author} {\bibfnamefont {A.}~\bibnamefont
  {Tajti}}, \bibinfo {author} {\bibfnamefont {P.~G.}\ \bibnamefont {Szalay}},
  \bibinfo {author} {\bibfnamefont {A.~G.}\ \bibnamefont {Császár}}, \bibinfo
  {author} {\bibfnamefont {M.}~\bibnamefont {Kállay}}, \bibinfo {author}
  {\bibfnamefont {J.}~\bibnamefont {Gauss}}, \bibinfo {author} {\bibfnamefont
  {E.~F.}\ \bibnamefont {Valeev}}, \bibinfo {author} {\bibfnamefont {B.~A.}\
  \bibnamefont {Flowers}}, \bibinfo {author} {\bibfnamefont {J.}~\bibnamefont
  {Vázquez}}, \ and\ \bibinfo {author} {\bibfnamefont {J.~F.}\ \bibnamefont
  {Stanton}},\ }\href {\doibase 10.1063/1.1811608} {\bibfield  {journal}
  {\bibinfo  {journal} {The Journal of Chemical Physics}\ }\textbf {\bibinfo
  {volume} {121}},\ \bibinfo {pages} {11599} (\bibinfo {year}
  {2004})}\BibitemShut {NoStop}%
\bibitem [{\citenamefont {Bak}\ \emph {et~al.}(2000)\citenamefont {Bak},
  \citenamefont {Jørgensen}, \citenamefont {Olsen}, \citenamefont {Helgaker},\
  and\ \citenamefont {Klopper}}]{Bak2000}%
  \BibitemOpen
  \bibfield  {author} {\bibinfo {author} {\bibfnamefont {K.~L.}\ \bibnamefont
  {Bak}}, \bibinfo {author} {\bibfnamefont {P.}~\bibnamefont {Jørgensen}},
  \bibinfo {author} {\bibfnamefont {J.}~\bibnamefont {Olsen}}, \bibinfo
  {author} {\bibfnamefont {T.}~\bibnamefont {Helgaker}}, \ and\ \bibinfo
  {author} {\bibfnamefont {W.}~\bibnamefont {Klopper}},\ }\href {\doibase
  10.1063/1.481544} {\bibfield  {journal} {\bibinfo  {journal} {The Journal of
  Chemical Physics}\ }\textbf {\bibinfo {volume} {112}},\ \bibinfo {pages}
  {9229} (\bibinfo {year} {2000})}\BibitemShut {NoStop}%
\bibitem [{\citenamefont {Hopkins}\ and\ \citenamefont
  {Tschumper}(2004)}]{Hopkins2004}%
  \BibitemOpen
  \bibfield  {author} {\bibinfo {author} {\bibfnamefont {B.~W.}\ \bibnamefont
  {Hopkins}}\ and\ \bibinfo {author} {\bibfnamefont {G.~S.}\ \bibnamefont
  {Tschumper}},\ }\href {\doibase 10.1021/jp0369084} {\bibfield  {journal}
  {\bibinfo  {journal} {The Journal of Physical Chemistry A}\ }\textbf
  {\bibinfo {volume} {108}},\ \bibinfo {pages} {2941} (\bibinfo {year}
  {2004})}\BibitemShut {NoStop}%
\bibitem [{\citenamefont {Bartlett}\ \emph {et~al.}(1990)\citenamefont
  {Bartlett}, \citenamefont {Watts}, \citenamefont {Kucharski},\ and\
  \citenamefont {Noga}}]{Bartlett1990}%
  \BibitemOpen
  \bibfield  {author} {\bibinfo {author} {\bibfnamefont {R.~J.}\ \bibnamefont
  {Bartlett}}, \bibinfo {author} {\bibfnamefont {J.}~\bibnamefont {Watts}},
  \bibinfo {author} {\bibfnamefont {S.}~\bibnamefont {Kucharski}}, \ and\
  \bibinfo {author} {\bibfnamefont {J.}~\bibnamefont {Noga}},\ }\href {\doibase
  10.1016/0009-2614(90)87031-L} {\bibfield  {journal} {\bibinfo  {journal}
  {Chemical Physics Letters}\ }\textbf {\bibinfo {volume} {165}},\ \bibinfo
  {pages} {513} (\bibinfo {year} {1990})}\BibitemShut {NoStop}%
\bibitem [{\citenamefont {Gyevi-Nagy}\ \emph {et~al.}(2019)\citenamefont
  {Gyevi-Nagy}, \citenamefont {Kállay},\ and\ \citenamefont
  {Nagy}}]{Gyevi2019}%
  \BibitemOpen
  \bibfield  {author} {\bibinfo {author} {\bibfnamefont {L.}~\bibnamefont
  {Gyevi-Nagy}}, \bibinfo {author} {\bibfnamefont {M.}~\bibnamefont {Kállay}},
  \ and\ \bibinfo {author} {\bibfnamefont {P.~R.}\ \bibnamefont {Nagy}},\
  }\href {\doibase 10.1021/acs.jctc.9b00957} {\bibfield  {journal} {\bibinfo
  {journal} {Journal of Chemical Theory and Computation}\ }\textbf {\bibinfo
  {volume} {16}},\ \bibinfo {pages} {366} (\bibinfo {year} {2019})}\BibitemShut
  {NoStop}%
\bibitem [{\citenamefont {Dunlap}\ \emph {et~al.}(1979)\citenamefont {Dunlap},
  \citenamefont {Connolly},\ and\ \citenamefont {Sabin}}]{Dunlap1979}%
  \BibitemOpen
  \bibfield  {author} {\bibinfo {author} {\bibfnamefont {B.~I.}\ \bibnamefont
  {Dunlap}}, \bibinfo {author} {\bibfnamefont {J.~W.~D.}\ \bibnamefont
  {Connolly}}, \ and\ \bibinfo {author} {\bibfnamefont {J.~R.}\ \bibnamefont
  {Sabin}},\ }\href {\doibase 10.1063/1.438728} {\bibfield  {journal} {\bibinfo
   {journal} {The Journal of Chemical Physics}\ }\textbf {\bibinfo {volume}
  {71}},\ \bibinfo {pages} {3396} (\bibinfo {year} {1979})}\BibitemShut
  {NoStop}%
\bibitem [{\citenamefont {Weigend}\ \emph {et~al.}(1998)\citenamefont
  {Weigend}, \citenamefont {Häser}, \citenamefont {Patzelt},\ and\
  \citenamefont {Ahlrichs}}]{Weigend1998}%
  \BibitemOpen
  \bibfield  {author} {\bibinfo {author} {\bibfnamefont {F.}~\bibnamefont
  {Weigend}}, \bibinfo {author} {\bibfnamefont {M.}~\bibnamefont {Häser}},
  \bibinfo {author} {\bibfnamefont {H.}~\bibnamefont {Patzelt}}, \ and\
  \bibinfo {author} {\bibfnamefont {R.}~\bibnamefont {Ahlrichs}},\ }\href
  {\doibase 10.1016/S0009-2614(98)00862-8} {\bibfield  {journal} {\bibinfo
  {journal} {Chemical Physics Letters}\ }\textbf {\bibinfo {volume} {294}},\
  \bibinfo {pages} {143} (\bibinfo {year} {1998})}\BibitemShut {NoStop}%
\bibitem [{\citenamefont {Røeggen}\ and\ \citenamefont
  {Johansen}(2008)}]{Roeggen2008}%
  \BibitemOpen
  \bibfield  {author} {\bibinfo {author} {\bibfnamefont {I.}~\bibnamefont
  {Røeggen}}\ and\ \bibinfo {author} {\bibfnamefont {T.}~\bibnamefont
  {Johansen}},\ }\href {\doibase 10.1063/1.2925269} {\bibfield  {journal}
  {\bibinfo  {journal} {The Journal of Chemical Physics}\ }\textbf {\bibinfo
  {volume} {128}},\ \bibinfo {pages} {194107} (\bibinfo {year}
  {2008})}\BibitemShut {NoStop}%
\bibitem [{\citenamefont {DePrince}\ and\ \citenamefont
  {Sherrill}(2013)}]{DePrince2013}%
  \BibitemOpen
  \bibfield  {author} {\bibinfo {author} {\bibfnamefont {A.~E.}\ \bibnamefont
  {DePrince}}\ and\ \bibinfo {author} {\bibfnamefont {C.~D.}\ \bibnamefont
  {Sherrill}},\ }\href {\doibase 10.1021/ct400250u} {\bibfield  {journal}
  {\bibinfo  {journal} {Journal of Chemical Theory and Computation}\ }\textbf
  {\bibinfo {volume} {9}},\ \bibinfo {pages} {2687} (\bibinfo {year}
  {2013})}\BibitemShut {NoStop}%
\bibitem [{\citenamefont {Häser}\ and\ \citenamefont
  {Almlöf}(1992)}]{Haser1992}%
  \BibitemOpen
  \bibfield  {author} {\bibinfo {author} {\bibfnamefont {M.}~\bibnamefont
  {Häser}}\ and\ \bibinfo {author} {\bibfnamefont {J.}~\bibnamefont
  {Almlöf}},\ }\href {\doibase 10.1063/1.462485} {\bibfield  {journal}
  {\bibinfo  {journal} {The Journal of Chemical Physics}\ }\textbf {\bibinfo
  {volume} {96}},\ \bibinfo {pages} {489} (\bibinfo {year} {1992})},\ \Eprint
  {http://arxiv.org/abs/https://doi.org/10.1063/1.462485}
  {https://doi.org/10.1063/1.462485} \BibitemShut {NoStop}%
\bibitem [{\citenamefont {Halko}\ \emph {et~al.}(2011)\citenamefont {Halko},
  \citenamefont {Martinsson},\ and\ \citenamefont {Tropp}}]{Halko2011}%
  \BibitemOpen
  \bibfield  {author} {\bibinfo {author} {\bibfnamefont {N.}~\bibnamefont
  {Halko}}, \bibinfo {author} {\bibfnamefont {P.~G.}\ \bibnamefont
  {Martinsson}}, \ and\ \bibinfo {author} {\bibfnamefont {J.~A.}\ \bibnamefont
  {Tropp}},\ }\href {\doibase 10.1137/090771806} {\bibfield  {journal}
  {\bibinfo  {journal} {SIAM Review}\ }\textbf {\bibinfo {volume} {53}},\
  \bibinfo {pages} {217} (\bibinfo {year} {2011})},\ \Eprint
  {http://arxiv.org/abs/https://doi.org/10.1137/090771806}
  {https://doi.org/10.1137/090771806} \BibitemShut {NoStop}%
\bibitem [{\citenamefont {Hohenstein}\ \emph
  {et~al.}(2012{\natexlab{a}})\citenamefont {Hohenstein}, \citenamefont
  {Parrish},\ and\ \citenamefont {Martínez}}]{Hohenstein2012}%
  \BibitemOpen
  \bibfield  {author} {\bibinfo {author} {\bibfnamefont {E.~G.}\ \bibnamefont
  {Hohenstein}}, \bibinfo {author} {\bibfnamefont {R.~M.}\ \bibnamefont
  {Parrish}}, \ and\ \bibinfo {author} {\bibfnamefont {T.~J.}\ \bibnamefont
  {Martínez}},\ }\href {\doibase 10.1063/1.4732310} {\bibfield  {journal}
  {\bibinfo  {journal} {The Journal of Chemical Physics}\ }\textbf {\bibinfo
  {volume} {137}},\ \bibinfo {pages} {044103} (\bibinfo {year}
  {2012}{\natexlab{a}})},\ \Eprint
  {http://arxiv.org/abs/https://doi.org/10.1063/1.4732310}
  {https://doi.org/10.1063/1.4732310} \BibitemShut {NoStop}%
\bibitem [{\citenamefont {Smith}\ \emph {et~al.}(2020)\citenamefont {Smith},
  \citenamefont {Burns}, \citenamefont {Simmonett}, \citenamefont {Parrish},
  \citenamefont {Schieber}, \citenamefont {Galvelis}, \citenamefont {Kraus},
  \citenamefont {Kruse}, \citenamefont {Remigio}, \citenamefont {Alenaizan},
  \citenamefont {James}, \citenamefont {Lehtola}, \citenamefont {Misiewicz},
  \citenamefont {Scheurer}, \citenamefont {Shaw}, \citenamefont {Schriber},
  \citenamefont {Xie}, \citenamefont {Glick}, \citenamefont {Sirianni},
  \citenamefont {O’Brien}, \citenamefont {Waldrop}, \citenamefont {Kumar},
  \citenamefont {Hohenstein}, \citenamefont {Pritchard}, \citenamefont
  {Brooks}, \citenamefont {Schaefer}, \citenamefont {Sokolov}, \citenamefont
  {Patkowski}, \citenamefont {DePrince}, \citenamefont {Bozkaya}, \citenamefont
  {King}, \citenamefont {Evangelista}, \citenamefont {Turney}, \citenamefont
  {Crawford},\ and\ \citenamefont {Sherrill}}]{Smith2020}%
  \BibitemOpen
  \bibfield  {author} {\bibinfo {author} {\bibfnamefont {D.~G.~A.}\
  \bibnamefont {Smith}}, \bibinfo {author} {\bibfnamefont {L.~A.}\ \bibnamefont
  {Burns}}, \bibinfo {author} {\bibfnamefont {A.~C.}\ \bibnamefont
  {Simmonett}}, \bibinfo {author} {\bibfnamefont {R.~M.}\ \bibnamefont
  {Parrish}}, \bibinfo {author} {\bibfnamefont {M.~C.}\ \bibnamefont
  {Schieber}}, \bibinfo {author} {\bibfnamefont {R.}~\bibnamefont {Galvelis}},
  \bibinfo {author} {\bibfnamefont {P.}~\bibnamefont {Kraus}}, \bibinfo
  {author} {\bibfnamefont {H.}~\bibnamefont {Kruse}}, \bibinfo {author}
  {\bibfnamefont {R.~D.}\ \bibnamefont {Remigio}}, \bibinfo {author}
  {\bibfnamefont {A.}~\bibnamefont {Alenaizan}}, \bibinfo {author}
  {\bibfnamefont {A.~M.}\ \bibnamefont {James}}, \bibinfo {author}
  {\bibfnamefont {S.}~\bibnamefont {Lehtola}}, \bibinfo {author} {\bibfnamefont
  {J.~P.}\ \bibnamefont {Misiewicz}}, \bibinfo {author} {\bibfnamefont
  {M.}~\bibnamefont {Scheurer}}, \bibinfo {author} {\bibfnamefont {R.~A.}\
  \bibnamefont {Shaw}}, \bibinfo {author} {\bibfnamefont {J.~B.}\ \bibnamefont
  {Schriber}}, \bibinfo {author} {\bibfnamefont {Y.}~\bibnamefont {Xie}},
  \bibinfo {author} {\bibfnamefont {Z.~L.}\ \bibnamefont {Glick}}, \bibinfo
  {author} {\bibfnamefont {D.~A.}\ \bibnamefont {Sirianni}}, \bibinfo {author}
  {\bibfnamefont {J.~S.}\ \bibnamefont {O’Brien}}, \bibinfo {author}
  {\bibfnamefont {J.~M.}\ \bibnamefont {Waldrop}}, \bibinfo {author}
  {\bibfnamefont {A.}~\bibnamefont {Kumar}}, \bibinfo {author} {\bibfnamefont
  {E.~G.}\ \bibnamefont {Hohenstein}}, \bibinfo {author} {\bibfnamefont
  {B.~P.}\ \bibnamefont {Pritchard}}, \bibinfo {author} {\bibfnamefont {B.~R.}\
  \bibnamefont {Brooks}}, \bibinfo {author} {\bibfnamefont {H.~F.}\
  \bibnamefont {Schaefer}}, \bibinfo {author} {\bibfnamefont {A.~Y.}\
  \bibnamefont {Sokolov}}, \bibinfo {author} {\bibfnamefont {K.}~\bibnamefont
  {Patkowski}}, \bibinfo {author} {\bibfnamefont {A.~E.}\ \bibnamefont
  {DePrince}}, \bibinfo {author} {\bibfnamefont {U.}~\bibnamefont {Bozkaya}},
  \bibinfo {author} {\bibfnamefont {R.~A.}\ \bibnamefont {King}}, \bibinfo
  {author} {\bibfnamefont {F.~A.}\ \bibnamefont {Evangelista}}, \bibinfo
  {author} {\bibfnamefont {J.~M.}\ \bibnamefont {Turney}}, \bibinfo {author}
  {\bibfnamefont {T.~D.}\ \bibnamefont {Crawford}}, \ and\ \bibinfo {author}
  {\bibfnamefont {C.~D.}\ \bibnamefont {Sherrill}},\ }\href {\doibase
  10.1063/5.0006002} {\bibfield  {journal} {\bibinfo  {journal} {The Journal of
  Chemical Physics}\ }\textbf {\bibinfo {volume} {152}},\ \bibinfo {pages}
  {184108} (\bibinfo {year} {2020})}\BibitemShut {NoStop}%
\bibitem [{\citenamefont {Wilke}\ \emph {et~al.}(2009)\citenamefont {Wilke},
  \citenamefont {Lind}, \citenamefont {Schaefer}, \citenamefont {Császár},\
  and\ \citenamefont {Allen}}]{Wilke2009}%
  \BibitemOpen
  \bibfield  {author} {\bibinfo {author} {\bibfnamefont {J.~J.}\ \bibnamefont
  {Wilke}}, \bibinfo {author} {\bibfnamefont {M.~C.}\ \bibnamefont {Lind}},
  \bibinfo {author} {\bibfnamefont {H.~F.}\ \bibnamefont {Schaefer}}, \bibinfo
  {author} {\bibfnamefont {A.~G.}\ \bibnamefont {Császár}}, \ and\ \bibinfo
  {author} {\bibfnamefont {W.~D.}\ \bibnamefont {Allen}},\ }\href {\doibase
  10.1021/ct900005c} {\bibfield  {journal} {\bibinfo  {journal} {Journal of
  Chemical Theory and Computation}\ }\textbf {\bibinfo {volume} {5}},\ \bibinfo
  {pages} {1511} (\bibinfo {year} {2009})}\BibitemShut {NoStop}%
\bibitem [{\citenamefont {Goerigk}\ and\ \citenamefont
  {Grimme}(2010)}]{Goerigk2010}%
  \BibitemOpen
  \bibfield  {author} {\bibinfo {author} {\bibfnamefont {L.}~\bibnamefont
  {Goerigk}}\ and\ \bibinfo {author} {\bibfnamefont {S.}~\bibnamefont
  {Grimme}},\ }\href {\doibase 10.1021/ct900489g} {\bibfield  {journal}
  {\bibinfo  {journal} {Journal of Chemical Theory and Computation}\ }\textbf
  {\bibinfo {volume} {6}},\ \bibinfo {pages} {107} (\bibinfo {year}
  {2010})}\BibitemShut {NoStop}%
\bibitem [{\citenamefont {Dunning}(1989)}]{Dunning1989}%
  \BibitemOpen
  \bibfield  {author} {\bibinfo {author} {\bibfnamefont {T.~H.}\ \bibnamefont
  {Dunning}},\ }\href {\doibase 10.1063/1.456153} {\bibfield  {journal}
  {\bibinfo  {journal} {The Journal of Chemical Physics}\ }\textbf {\bibinfo
  {volume} {90}},\ \bibinfo {pages} {1007} (\bibinfo {year}
  {1989})}\BibitemShut {NoStop}%
\bibitem [{\citenamefont {Woon}\ and\ \citenamefont
  {Dunning}(1993)}]{Woon1993}%
  \BibitemOpen
  \bibfield  {author} {\bibinfo {author} {\bibfnamefont {D.~E.}\ \bibnamefont
  {Woon}}\ and\ \bibinfo {author} {\bibfnamefont {T.~H.}\ \bibnamefont
  {Dunning}},\ }\href {\doibase 10.1063/1.464303} {\bibfield  {journal}
  {\bibinfo  {journal} {The Journal of Chemical Physics}\ }\textbf {\bibinfo
  {volume} {98}},\ \bibinfo {pages} {1358} (\bibinfo {year}
  {1993})}\BibitemShut {NoStop}%
\bibitem [{\citenamefont {Woon}\ and\ \citenamefont
  {Dunning}(1994)}]{Woon1994}%
  \BibitemOpen
  \bibfield  {author} {\bibinfo {author} {\bibfnamefont {D.~E.}\ \bibnamefont
  {Woon}}\ and\ \bibinfo {author} {\bibfnamefont {T.~H.}\ \bibnamefont
  {Dunning}},\ }\href {\doibase 10.1063/1.466439} {\bibfield  {journal}
  {\bibinfo  {journal} {The Journal of Chemical Physics}\ }\textbf {\bibinfo
  {volume} {100}},\ \bibinfo {pages} {2975} (\bibinfo {year}
  {1994})}\BibitemShut {NoStop}%
\bibitem [{\citenamefont {Papajak}\ and\ \citenamefont
  {Truhlar}(2011)}]{Papajak2011}%
  \BibitemOpen
  \bibfield  {author} {\bibinfo {author} {\bibfnamefont {E.}~\bibnamefont
  {Papajak}}\ and\ \bibinfo {author} {\bibfnamefont {D.~G.}\ \bibnamefont
  {Truhlar}},\ }\href {\doibase 10.1021/ct1005533} {\bibfield  {journal}
  {\bibinfo  {journal} {Journal of Chemical Theory and Computation}\ }\textbf
  {\bibinfo {volume} {7}},\ \bibinfo {pages} {10} (\bibinfo {year}
  {2011})}\BibitemShut {NoStop}%
\bibitem [{\citenamefont {Jurečka}\ \emph {et~al.}(2006)\citenamefont
  {Jurečka}, \citenamefont {Šponer}, \citenamefont {Černý},\ and\
  \citenamefont {Hobza}}]{Jureka2006}%
  \BibitemOpen
  \bibfield  {author} {\bibinfo {author} {\bibfnamefont {P.}~\bibnamefont
  {Jurečka}}, \bibinfo {author} {\bibfnamefont {J.}~\bibnamefont {Šponer}},
  \bibinfo {author} {\bibfnamefont {J.}~\bibnamefont {Černý}}, \ and\
  \bibinfo {author} {\bibfnamefont {P.}~\bibnamefont {Hobza}},\ }\href
  {\doibase 10.1039/B600027D} {\bibfield  {journal} {\bibinfo  {journal} {Phys.
  Chem. Chem. Phys.}\ }\textbf {\bibinfo {volume} {8}},\ \bibinfo {pages}
  {1985} (\bibinfo {year} {2006})}\BibitemShut {NoStop}%
\bibitem [{\citenamefont {Gráfová}\ \emph {et~al.}(2010)\citenamefont
  {Gráfová}, \citenamefont {Pitoňák}, \citenamefont {Řezáč},\ and\
  \citenamefont {Hobza}}]{Grafova2010}%
  \BibitemOpen
  \bibfield  {author} {\bibinfo {author} {\bibfnamefont {L.}~\bibnamefont
  {Gráfová}}, \bibinfo {author} {\bibfnamefont {M.}~\bibnamefont
  {Pitoňák}}, \bibinfo {author} {\bibfnamefont {J.}~\bibnamefont {Řezáč}},
  \ and\ \bibinfo {author} {\bibfnamefont {P.}~\bibnamefont {Hobza}},\ }\href
  {\doibase 10.1021/ct1002253} {\bibfield  {journal} {\bibinfo  {journal}
  {Journal of Chemical Theory and Computation}\ }\textbf {\bibinfo {volume}
  {6}},\ \bibinfo {pages} {2365} (\bibinfo {year} {2010})}\BibitemShut
  {NoStop}%
\bibitem [{\citenamefont {Parrish}\ \emph {et~al.}(2012)\citenamefont
  {Parrish}, \citenamefont {Hohenstein}, \citenamefont {Martínez},\ and\
  \citenamefont {Sherrill}}]{Parrish2012}%
  \BibitemOpen
  \bibfield  {author} {\bibinfo {author} {\bibfnamefont {R.~M.}\ \bibnamefont
  {Parrish}}, \bibinfo {author} {\bibfnamefont {E.~G.}\ \bibnamefont
  {Hohenstein}}, \bibinfo {author} {\bibfnamefont {T.~J.}\ \bibnamefont
  {Martínez}}, \ and\ \bibinfo {author} {\bibfnamefont {C.~D.}\ \bibnamefont
  {Sherrill}},\ }\href {\doibase 10.1063/1.4768233} {\bibfield  {journal}
  {\bibinfo  {journal} {The Journal of Chemical Physics}\ }\textbf {\bibinfo
  {volume} {137}},\ \bibinfo {pages} {224106} (\bibinfo {year}
  {2012})}\BibitemShut {NoStop}%
\bibitem [{\citenamefont {Hohenstein}\ \emph
  {et~al.}(2012{\natexlab{b}})\citenamefont {Hohenstein}, \citenamefont
  {Parrish}, \citenamefont {Sherrill},\ and\ \citenamefont
  {Martínez}}]{Hohenstein2012b}%
  \BibitemOpen
  \bibfield  {author} {\bibinfo {author} {\bibfnamefont {E.~G.}\ \bibnamefont
  {Hohenstein}}, \bibinfo {author} {\bibfnamefont {R.~M.}\ \bibnamefont
  {Parrish}}, \bibinfo {author} {\bibfnamefont {C.~D.}\ \bibnamefont
  {Sherrill}}, \ and\ \bibinfo {author} {\bibfnamefont {T.~J.}\ \bibnamefont
  {Martínez}},\ }\href {\doibase 10.1063/1.4768241} {\bibfield  {journal}
  {\bibinfo  {journal} {The Journal of Chemical Physics}\ }\textbf {\bibinfo
  {volume} {137}},\ \bibinfo {pages} {221101} (\bibinfo {year}
  {2012}{\natexlab{b}})}\BibitemShut {NoStop}%
\end{thebibliography}%

\end{document}